# Experimentally shock-induced melt veins in basalt: Improving the shock classification of eucrites

Haruka Ono[1], Kosuke Kurosawa[1,*], Takafumi Niihara[2], Takashi Mikouchi[3], Naotaka Tomioka[4], Junko Isa[1], Hiroyuki Kagi[5], Takuya Matsuzaki[6], Hiroshi Sakuma[7], Hidenori Genda[8], Tatsuhiro Sakaiya[9], Tadashi Kondo[9], Masahiro Kayama[10], Mizuho Koike[11], Yuji Sano[6], Masafumi Murayama[6], Wataru Satake[12] and Takafumi Matsui[1,12]

[1]Planetary Exploration Research Center, Chiba Institute of Technology, 2-17-1, Tsudanuma, Narashino, Chiba 275-0016, Japan

[2]Department of Applied Science, Okayama University of Science, 1-1 Ridaicho, Kita-ku, Okayama-shi 700-0005, Japan

[3]The University Museum, The University of Tokyo, 7-3-1 Hongo, Bunkyo-ku, Tokyo 113-0033, Japan

[4]Kochi Institute for Core Sample Research, X-star, Japan Agency for Marine-Earth Science and Technology (JAMSTEC), 200 Monobe Otsu, Nankoku, Kochi 783-8502, Japan

[5]Geochemical Research Center, Graduate School of Science, The University of Tokyo, 7-3-1 Hongo, Bunkyo-ku, Tokyo 113-0033, Japan

[6]Center for Advanced Marine Core Research, Kochi University, 200 Monobe Otsu, Nankoku, Kochi 783-8502, Japan

[7] Research Center for Functional Materials, National Institute for Materials Science, 1-1, Namiki, Tsukuba, Ibaraki 305-0044, Japan

[8]Earth–Life Science Institute, Tokyo Institute of Technology, 2-12-1 Ookayama, Meguro-ku, Tokyo 152-8550, Japan

[9]Department of Earth and Space Science, Graduate School of Science, Osaka University, 1-1 Machikaneyama, Toyonaka, Osaka 560-0043, Japan

[10]Department of General Systems Studies, Graduate School of Arts and Sciences, The Univ. of Tokyo, 3-8-1, Komaba, Meguro-ku, Tokyo 153-8902, Japan

[11] Earth and Planetary Systems Science Program, Graduate School of Advanced Science and Engineering, Hiroshima University, 1-3-1, Kagamiyama, Higashi–Hiroshima, Hiroshima 739-8526, Japan

[12]Institute of Geo-Cosmology, Chiba Institute of Technology, 2-17-1, Tsudanuma, Narashino, Chiba 275-0016, Japan

[*]Corresponding author: Kosuke Kurosawa (kosuke.kurosawa@perc.it-chiba.ac.jp)

**Abstract**

Basaltic rocks occur widely on the terrestrial planets and differentiated asteroids, including the asteroid 4 Vesta. We conducted a shock recovery experiment with decaying compressive pulses on a terrestrial basalt at Chiba Institute of Technology, Japan. The sample recorded a range of pressures, and shock physics modeling was conducted to add a pressure scale to the observed shock features. The shocked sample was examined by optical and electron microscopy, electron back-scattered diffractometry, and Raman spectroscopy. We found that localized melting occurs at a lower pressure (~10 GPa) than previously thought (>20 GPa). The shocked basalt near the epicenter represents "shock degree C" of a recently proposed classification scheme for basaltic eucrites and, as such, our results provide a pressure scale for the classification scheme. Finally, we estimated the total fraction of the basaltic eucrites classified as shock degree C to be ~15% by assuming the impact velocity distribution onto Vesta.

**1. Introduction**

Shock metamorphic features in meteorites are evidence of ancient impact events, which were one of the major processes involved in the evolution of the solar system [e.g., Stöffler et al., 1991, 2018]. Therefore, it is possible to constrain the nature of impact events, such as the speed of mutual collisions, by obtaining information from impact metamorphic textures. In this study, we focused on eucrites, which are achondritic meteorites, because a number of eucrites with shock features have been found [e.g., McSween et al., 2011; Kanemaru et al., 2020]. In addition, it is thought that eucrites originate from the crust of the asteroid Vesta [e.g., Mittlefehldt, 2015], which is the second largest asteroid in the solar system. If we could obtain an accurate understanding of the relationship between the degree of shock metamorphism and impact conditions, it would be possible to constrain the impact environment of Vesta and impact conditions in eucrites. The dynamical circumstances relevant to other large planetesimals or proto-planets might be similar to Vesta.

Shock metamorphism in eucrites has been investigated in previous studies [e.g., Yamaguchi et al., 1997; Miyahara et al., 2014]. Recently, Kanemaru et al. (2020) investigated 12 basaltic and 4 cumulate eucrites systematically, and classified the shock features in eucrites into progressive shock degrees from A to E. Hereafter, we refer to this shock degree classification as the "Kanemaru table". Stöffler et al. (1991, 2018) also provided a shock classification scheme, which has been referred to as the "Stöffler table". We summarize these two tables in Table 1. A weak point of the Kanemaru table is that the peak pressures required to produce the shock features are unknown. The five shock degrees determined by Kanemaru et al. (2020) correspond to the shock stages S1–S3 of the Stöffler table, suggesting that a finer classification of shock degree can be made if the Kanemaru table were to be linked to the peak pressures.

In this study, we conducted shock recovery experiments with terrestrial basalt blocks using a recently developed technique [Kurosawa et al., 2022], which allows collection of shocked samples that have experienced a range of peak pressures and temperatures. The position-dependent peak pressures and temperatures were estimated by shock physics modeling. We identified the shocked region corresponding to shock degree

"C" in the Kanemaru table, and determined the peak pressure of this region to be 10–20 GPa. We then discuss the impact processes on Vesta based on the Kanemaru table.

**Table 1.** Shock stages and classification [Stöffler et al., 2018; Kanemaru et al., 2020]. The symbols "-" and "+" mean "not identified" and "identified", respectively. Note that the original version of "Table M" in Stöffler et al. (2018) did not mention undulatory extinction in major silicates in mafic rocks. As such, we added undulatory extinction to shock stage S2, based on the required shock pressures for producing undulatory extinction in plagioclase, olivine, and pyroxene.

| Modified after Table M of Stöffler et al. (2018) | | | | Kanemaru et al. (2020) | | | | |
|---|---|---|---|---|---|---|---|---|
| Shock stage | Shock pressure (GPa) | Shock effect in major minerals | Accompanying disequilibrium shock effect | Shock degree | Shock pressure (GPa) | Undulatory extinction angles | Abundance of maskelynite | Presence of shock veins |
| S1 | ~1–5 | Sharp optical extinction of all minerals. Irregular fracturing. | - | A | ~10[d] | <2° | - | - |
| S2 | | Fractured silicates; mechanical twinning of pyroxene; kink bands in mica; undulatory extinction in silicates[a]. | - | B | | ≤15° | - | - |
| S3 | ~20–22 | Plagioclase with pdf[b] and partially converted to diaplectic glass; mosaicism in plagioclase and mafic minerals; pf[c] in olivine. | Incipient formation of localized "mixed melt" and glassy melt veins. | C | | >15° | - | + |
| S4 | ~28–34 | Diaplectic plagioclase glass; mechanical twinning in pyroxene; mosaicism in mafic minerals; pf in olivine. | Localized "mixed melt" and melt veins (glassy or microcrystalline). | D | ~20[e] | >15° | <10% | + |
| S5 | ~42–45 | Melted plagioclase glass with incipient flow structure and vesicles; mosaicism in mafic minerals; pdf in pyroxene and amphibole; loss of pleochroism in mafic minerals. | Melt veins and pockets. | E | | | >50% | + |
| S6 | ~55–60 | Melted plagioclase glass with vesicles and flow structure; incipient and increasing contact melting of pyroxene and olivine, and (re)crystallization of olivine. | Pervasive melt veins and melt pockets. | | | | | |
| S7 | >60–65 | Complete melting | | | | | | |

[a] Based on shock effects in plagioclase, pyroxene, and olivine (figures 14, 16, and 17 in Stöffler et al., 2018).

[b] Planar fracture

[c] Planar deformation feature

## 2. Materials and Methods

### 2.1. Impact experiment

A shock recovery experiment was conducted with a two-stage light gas gun at the Planetary Exploration Research Center of the Chiba Institute of Technology, Japan [Kurosawa et al., 2015]. The experimental method employed in this study is basically the same as that developed by Kurosawa et al. (2022), except for the samples that were used. The method allows us to collect shocked samples that retain the pre-impact stratigraphy. We used a terrestrial basalt from Inner Mongolia as a target. The reference density $\rho_0$ of the basalt (2.94 Mg m$^{-3}$) is comparable to that of known eucrites (2.66 – 3.12 Mg m$^{-3}$) [Macke et al., 2011]. Given that there are no open pores or cracks with a size larger than ~1 μm, we neglected the effects of porosity compaction on the impact outcomes. The detailed sample description is presented in the Supplementary Information (Text S1; Figures S1–S6; Tables S1–S2). The basalt samples were shaped into cylinders with a diameter of 30 mm and height of 24 mm. We conducted three shots (#471, #476, and #492). The samples were sealed in a titanium (Ti) container and covered with a detachable Ti front plate with a thickness of 4 mm for shot #471 and 3 mm for shots #476 and #492. The impact velocities were 7.3 km s$^{-1}$ (#471), 7.0 km s$^{-1}$ (#476), and 6.9 km s$^{-1}$ (#492). The projectiles were polycarbonate spheres with a diameter of 4.8 mm (shots #471 and #476) and a titanium sphere with a diameter of 2.0 mm (shot #492). The reference densities of the polycarbonate and titanium are 1.12 Mg m$^{-3}$ and 4.51 Mg m$^{-3}$, respectively.

### 2.2. Sample analyses

We analyzed polished thin-sections made from the cross-sections of the entire recovered samples across the impact-generated depressions on the shocked samples. We examined these with optical microscopy, scanning electron microscopy (SEM; JEOL JSM-6510LA at the Chiba Institute of Technology), and field emission SEM (FE-SEM; JEOL JSM-7000F at the University of Tokyo) methods. Chemical analyses of plagioclase, pyroxene, olivine, and opaque minerals in the thin-sections were performed with two electron probe microanalyzers (JEOL JXA-8530F and JXA-8900L at the University of Tokyo).

We investigated whether the shocked samples are crystalline or quenched glass using an electron back-scattered diffraction (EBSD) methods with the FE-SEM. In addition, we conducted micro-Raman analyses at the University of Tokyo to investigate the shock-generated glass, which formed immediately after the impact. Further details are described in the Supplementary Information (Text S2).

We also investigated the damage distribution in the shocked sample from shot #476 by micro-focus X-ray computational tomography (XCT; Zeiss Xradia 410 Versa) at the Center for Advanced Marine Core Research, Kochi University, Japan.

## 2.3. Shock physics modeling

We conducted numerical calculations using the same impact conditions as the experiments using the two-dimensional version of the iSALE shock physics code [Amsden et al., 1981; Ivanov et al., 1997; Wünnemann et al., 2006; Collins et al., 2016], in order to estimate the peak pressure and peak temperature distributions in the shocked samples. The model settings and material model parameters were basically the same as used in Kurosawa et al. (2022), except that the target material was basalt used in the present study. As such, we used the ANalytical Equations of State (ANEOS basalt) [Thompson and Lauson, 1972; Pierazzo et al., 2005; Sato et al., 2021] and rock strength model [Collins et al., 2004] for basalt [Bowling et al., 2020; Ivanov et al., 2010] in this study. Further details can be found in the Supplementary Information (Text S3; Tables S2–S3).

## 3. Results

### 3.1. Peak pressure and temperature distributions

Figure 1a shows the peak pressure and temperature distributions obtained from the impact simulation for shot #471. The peak pressure at the epicenter in the basalt, which is immediately beneath the impact point on the Ti plate, is 14 GPa. The peak pressure decreases gradually with increasing distance down to ~1 GPa. The estimated maximum temperature at the epicenter was ~900 K. The temperature increase is negligible at >10 mm from the epicenter. We investigated the validity of the iSALE results by comparing these with the experimental results of Nakazawa et al. (2002), and confirmed that the iSALE results are robust, particularly in the region with peak pressures over 10 GPa. Further details are described in the Supplementary Information (Text S4; Figures S7–S8; Tables S4–S5).

### 3.2. General petrology

We confirmed that the damage structure in the shocked basalt has no clear dependence on the azimuth angle with respect to the impact direction, based on the 3D-XCT results for shot #476 (Text S5; Figure S9). Hereafter, we only present the results from the observations of polished thin-sections, which were cut parallel to the impact direction. Figure 1b shows a mosaic of back-scattered electron (BSE) images. The recovered samples contain many open cracks, and silicates in these samples exhibit stronger undulatory extinction than in the intact samples under a polarizing microscope. (Texts S6–S7; Figure S10; Movies S1–S3). However, shock features due to greater shock pressures, including planar deformation features (PDFs), mosaicism, and maskelynite formation in plagioclase, were not produced even around the impact epicenter (14 GPa for shot #471 and 19 GPa for shot #492).

We identified six glass veins within a distance of 10 mm of the epicenter (Figure 1b). The veins do not exhibit any birefringence under cross-polarized light, indicating that the veins are quenched melts, which have frequently been referred to as shock melt veins (SMVs) in meteorites.

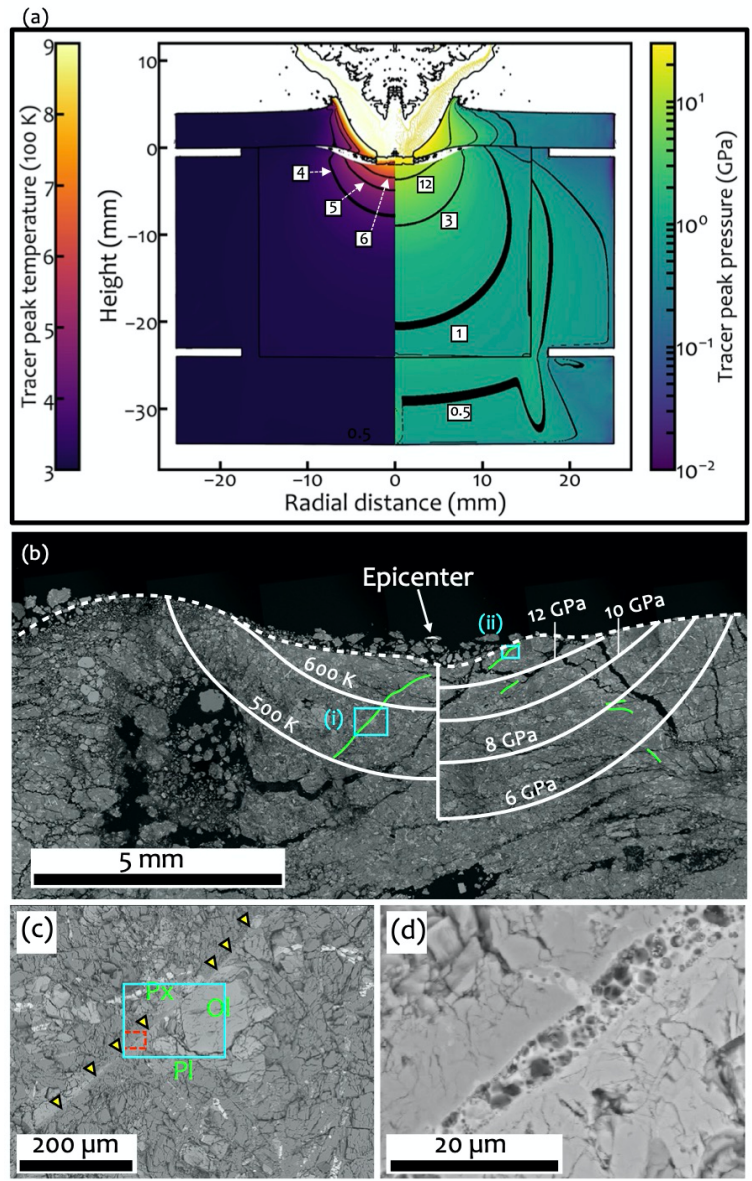

**Figure 1.** (a) Peak temperature (left) and peak pressure (right) distributions calculated with iSALE at a vertical impact velocity of 7.3 km s$^{-1}$. The isolines of pressures and temperatures are shown as black lines. (b) Enlarged back-scattered electron (BSE) image of the recovered sample around the epicenter (shot #471). The isobaric and isothermal lines calculated with iSALE are shown as white lines. The green lines indicate the locations of the shock melt veins (SMVs). (c) Enlarged BSE image of the cyan-solid box (i) shown in (b). Yellow triangles indicate the shock melt vein (Px = pyroxene, Ol = olivine, and Pl = plagioclase). (d) Enlarged BSE image of the red-dashed box in (c), which shows the presence of tiny vesicles in the SMV.

### 3.3. Detailed observations of the shock melt veins

In this section, we describe in detail the SMVs in our shocked sample. Figure 1c–d show a close-up of the identified SMVs. The SMVs are 1–4 µm in width and contain tiny vesicles. The diameter of the vesicles ranges from 0.1 to 2.0 µm, although smaller vesicles cannot be observed due to the limitations of spatial resolution. The vesicles may be voids due to devolatilization of the sample. We obtained an EBSD pattern and a Raman spectrum for a location inside a SMV (point A in Figure 2b) to investigate whether the materials in the SMV were melted or not. For comparison, locations outside of the SMV, which appear to be crystalline, were also analyzed by the EBSD (point B in Figure 2b) and Raman (point C in Figure 2b) methods. Both the EBSD pattern (Figure 2c) and Raman spectrum (Figure 2e) of point A do not exhibit any crystalline features (i.e., we did not obtain a clear Kikuchi line in the EBSD data and Raman peaks of crystalline silicates), which shows that the materials in the SMV were melted due to localized heating above the basalt solidus. In contrast, outside the SMV the materials are still crystalline (Figure 2d and f). To constrain the peak pressure experienced by the SMVs, we investigated plagioclase adjacent to the SMVs with an optical microscope. We confirmed that the plagioclase exhibited stronger undulatory extinction than the starting material. However, plagioclase conversion to maskelynite was not observed. Consequently, the materials in the SMVs were heated above the solidus, but are classified as shock stage S2 according to the Stöffler table.

The chemical compositions of the SMVs were measured with the electron probe microanalyzer. We compared these with the compositions of plagioclase, pyroxene, and olivine in the intact basalt using normalized oxide compositions (i.e., $Al_2O_3/SiO_2$ versus $MgO/SiO_2$ and $CaO/SiO_2$ versus $FeO/SiO_2$; Figure 3). These diagrams suggest that the materials in the SMVs are mixtures of plagioclase and olivine, because an olivine phenocryst was located near the region (Figure 1c). An elemental map around a SMV located in the region within the blue box (ii) in Figure 1b is presented in the Supplementary Information (Text S8; Figure S11).

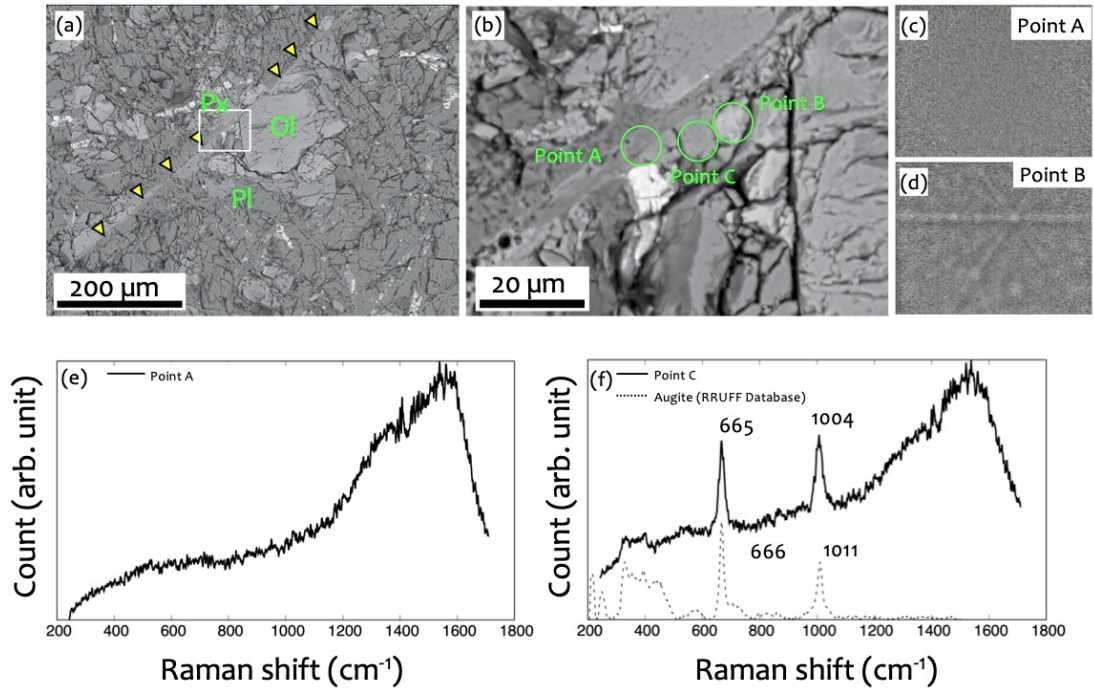

**Figure 2.** (a) BSE image of a SMV (same as Figure 1c). (b) Enlarged BSE image of the white box in (a). The measurement points A–C for the EBSD and Raman analyses are shown. (c) Observed EBSD pattern at point A inside a SMV. (d) Observed EBSD pattern at point B outside a SMV. (e) Raman spectrum at point A. (f) Raman spectrum at point C. The reference spectrum for augite was taken from the RRUFF database [Lafuente et al., 2015], and is shown as a dashed line.

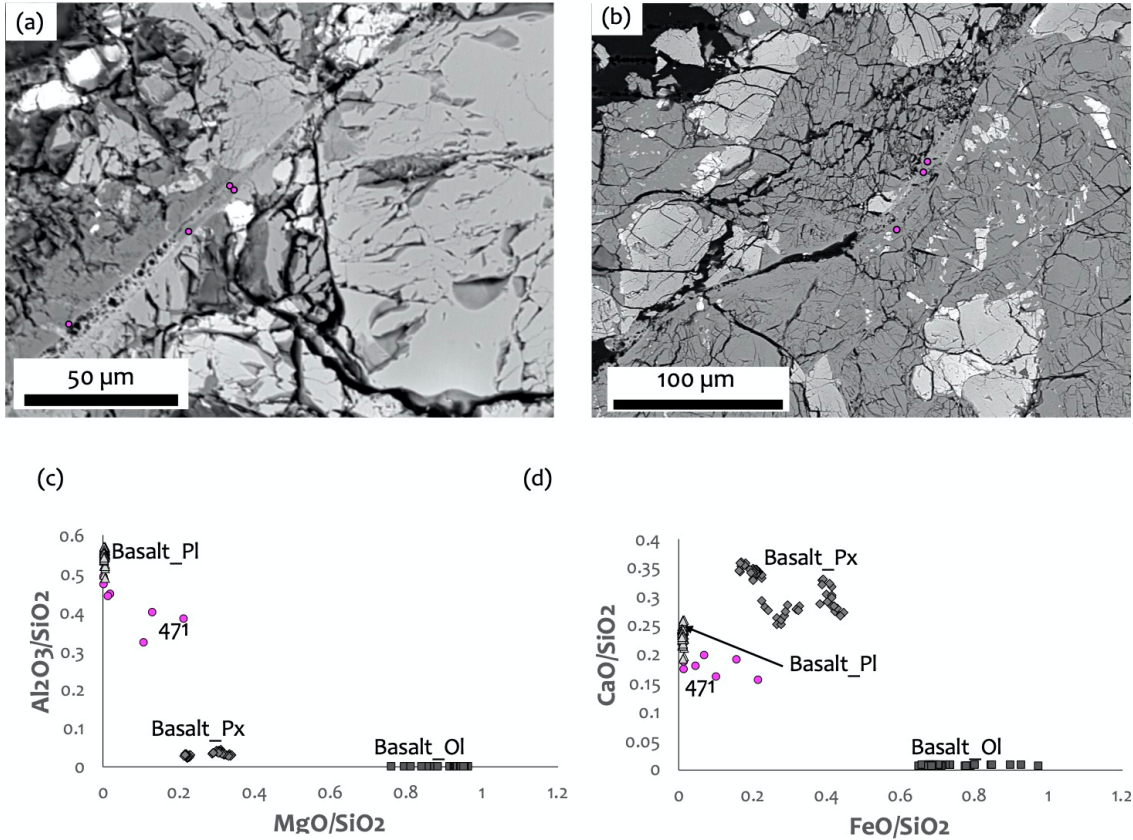

**Figure 3.** (a–b) Locations of the electron probe microanalyses shown as dots in magenta on enlarged BSE images. The locations shown in (a) and (b) correspond to the cyan rectangle in Figure 1c and cyan rectangle in Figure 1b-(ii), respectively. (c) $Al_2O_3/SiO_2$ versus $MgO/SiO_2$. (d) $CaO/SiO_2$ versus $FeO/SiO_2$. The data are for shot #471 and intact basalt (Pl = plagioclase, Px = pyroxene, and Ol = olivine).

## 4. Discussion

### 4.1. Connection between the two different shock stage classification schemes

Our optical microscopy observations suggest that the region around the SMVs in the recovered sample represent shock stage S2 of the Stöffler table, based on the lack of maskelynite (section 3.2). The estimated peak pressure with the iSALE pertaining to this region is ~10 GPa (Figure 1b), which is consistent with the required shock pressure for producing the shock features of S2. However, the presence of SMVs is one of the features of shock stage S3 (20–30 GPa) in the Stöffler table (Table 1). Consequently, our results lead to a downward revision of the required peak pressure for producing SMVs from 20 to 10 GPa. The undulatory extinction of silicates, presence of the SMVs, and absence of maskelynite are consistent with shock degree C in the Kanemaru table (Table 1). Although we also identified a SMV in the region where the peak pressure was ~6 GPa (Figure 1b), the SMV is rather short and narrow. As such, we conservatively propose that the required peak pressure for producing shocked basaltic rocks with a shock degree C is 10 GPa. We also identified SMVs within the ~10 GPa region in the thin-section of shot #492 (Text S9; Figure S12).

Our results allow a pressure scale to be applied to the Kanemaru table. These results and the Stöffler table suggest that shock degree C of the Kanemaru table corresponds to 10–20 GPa for basaltic rocks. Shock degrees A and B represent <10 GPa compression, and basaltic eucrites classified into shock degrees D and E underwent >20 GPa compression.

**4.2. Localized heating in basaltic rocks**

The iSALE simulations also provide the peak temperature with distance from the impact epicenter. The estimated peak temperature at the epicenter is ~900 K, which is ~300 K lower than the solidus for basaltic rocks [Ivanov et al., 2010]. In addition, we found an SMV at a location where the estimated peak temperature was estimated to be ~500 K (Figure 1a). The effect of plastic deformation on heating [Ivanov et al., 2002; Quintana et al., 2015; Kurosawa and Genda, 2018; Kurosawa, Genda et al., 2021] was taken into account in the iSALE simulations, and the effect of porosity compaction on heating can be neglected (Text S1). Consequently, local energy concentration, which is not accounted for by iSALE, would have occurred in the shocked basalt. The local heating may be caused by shear velocities between minerals with different densities [e.g., Kurosawa, Moriwaki et al., 2021], because our basalt sample contains several different minerals (Text S1). Nevertheless, the actual mechanism that produces the SMVs should be investigated further.

**4.3. Implications for impact processes on Vesta**

We conducted another series of iSALE simulations that consider impacts onto the asteroid Vesta. Hereafter, we only discuss the results of this Vesta model. The details of the simulations are presented in the Supplementary Information (Text S10; Table S6). The key differences from the simulation described in Section 2.3 are briefly described as follows. The material model for the projectile and Vestian surface were the same as used for the basalt samples (section 2.3). We considered vertical impacts only. The impact velocity was varied from 2–14 km s$^{-1}$ based on the impact velocity distribution on Vesta [Schmedemann et al., 2014]. We assumed that the target body has a uniform temperature of 220 K, which is close to the equilibrium temperature at the current orbit of Vesta. Figure 4a shows the initial locations of the shocked materials classified as shock degree C (10–20 GPa) at different impact velocities. If we consider an average impact velocity onto Vesta (4.6 km s$^{-1}$), the maximum depths of the 10 GPa compression are close to the projectile diameter. For comparison, we also show the result of an impact at 10 km s$^{-1}$ in the figure. The maximum depth corresponding to 10 GPa extends down to three projectile diameters. Although the maximum depth of a given peak pressure becomes somewhat shallower in the case of oblique impacts, the effects of the impact obliquity on the maximum depth is not significant unless the impact angle is <30° as measured from the target surface [Pierazzo and Melosh, 2000]. Figure 4b shows the shocked target masses $M_C$ and $M_{D\&E}$, which correspond to the masses of the shocked materials classified into shock degree C (10–20 GPa) and shock degrees D and E (>20 GPa), as a function of the impact velocity. We fitted $M_C$ and $M_{D\&E}$ with a third-order polynomial function as follows:

$$\log_{10}(M_i) = a_0 + a_1 v_{imp} + a_2 v_{imp}^2 + a_3 v_{imp}^3, \quad (1)$$

where $M_i$ is the shocked mass normalized to the projectile mass (i = C or D&E), $a_0$ to $a_3$ are the fitting constants, and $v_{imp}$ is the impact velocity. $a_0$ to $a_3$ for $M_C$ and $M_{D\&E}$ are $a_0 = -3.1096$, $a_1 = 1.1251$, $a_2 = -0.096164$, and $a_3 = 0.0028883$, and $a_0 = -4.5908$, $a_1 = 1.3355$, $a_2 = -0.10407$, and $a_3 = 0.0028581$, respectively. The impact probability onto Vesta is also shown in the figure. Given that we did not use a scale-dependent strength model and neglected gravity, the results can be applied to any spatial scale if the target curvature can be neglected. We estimated the velocity-weighted average masses of $M_C$ and $M_{D\&E}$ to be 1.9 and 0.65 projectile masses, respectively, by convolving the $v_{imp}$ distribution.

Rubin (2015) showed that the number fraction of basaltic eucrites that are maskelynite-bearing, i.e., classified into the shock degree D or E, is only ~5%. This number is the outcome of all the processes experienced by basaltic eucrites, including shock metamorphism on Vesta, impact gardening on the Vestian surface, launch from Vesta due to the latest impact event, and entry into Earth's atmosphere. Our experimental results predict that the number fraction of basaltic eucrites classified into shock degree C of the Kanemaru table is ~15%, although such verification is beyond the scope of this study. If the fraction deviates significantly from this number, there are two possible explanations: (1) Selective process(es), such as selective escape from Vesta depending on ejection velocity during the impact events [e.g., Güldemeister et al. 2022], might more likely cause shocked eucrites that do not contain maskelynite to be classified as shock degree C than Shock degrees D and E; and (2) the velocity distribution onto Vesta may have changed at some time, because this was used for that currently on Vesta to obtain the averaged $M_C$ and $M_{D\&E}$.

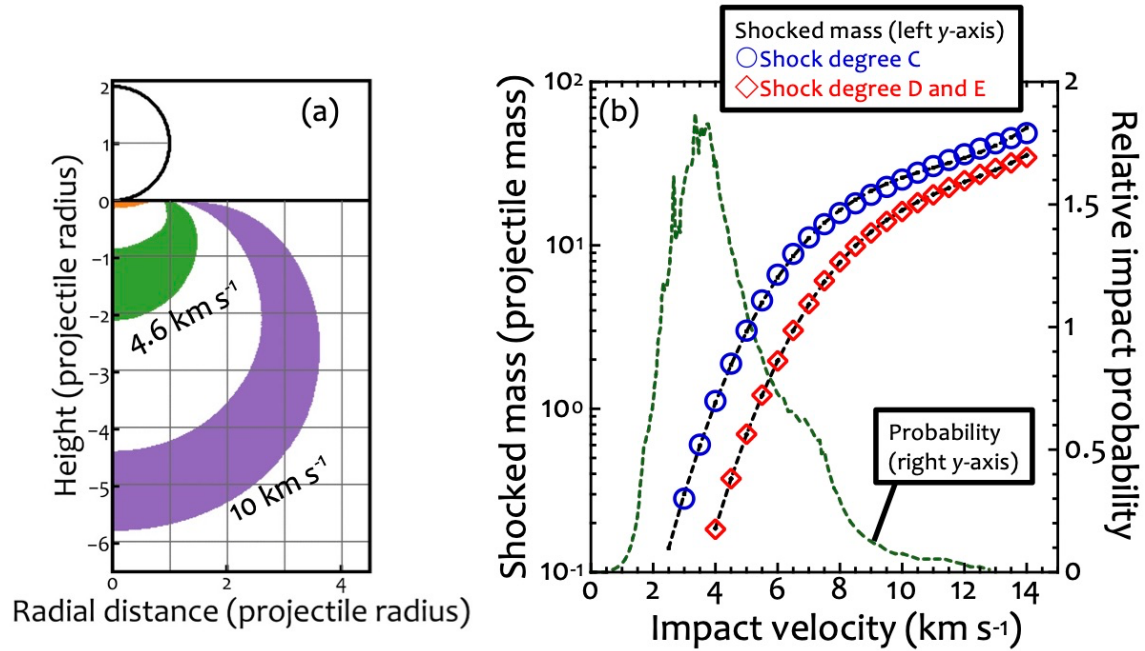

**Figure 4.** iSALE results. These simulations were conducted to estimate the shocked mass on Vesta (Section 4.3; Supplementary Text S10). (a) Initial locations of the shocked materials experiencing 10–20 GPa compression. The results are shown for three different impact velocities (orange = 2 km s$^{-1}$, green = 4.6 km s$^{-1}$, and purple = 10 km s$^{-1}$). The gray grid has a spacing that corresponds to the projectile radius. (b) The shocked mass that corresponds to shock degree C (blue circles; $M_C$) and shock degree D and E (red squares; $M_{D\&E}$) as a function of the impact velocity. The black dashed lines are the best-fit curves defined by Eq. (1). The impact velocity distribution onto Vesta [Schmedemann et al., 2014] (green dashed line) is also shown on the right $y$-axis.

## 5. Conclusions

We undertook a shock recovery experiment on a terrestrial basalt sample to investigate the shock effects in extraterrestrial basaltic rocks, such as eucrites. Our microscopic observations and shock physics modeling confirmed that shock melt veins are produced at pressures of >10 GPa. These conditions for localized melting are lower than previously thought (20 GPa). The shock features of materials near the impact epicenter are consistent with shock degree C of the classification scheme proposed by Kanemaru et al. (2020). Based on the presence of SMVs and lack of maskelynite, we propose that shock degree C corresponds to 10–20 GPa for basaltic rocks. Finally, we predict that the fraction of basaltic eucrites classified as shock degree C is ~15%, which is about three times greater than the fraction of maskelynite-bearing basaltic eucrites.

**Open Research Statement**

We used PDXL software (Rigaku, https://www.rigaku.com/support/software/pdxl), which is also a commercial package, to identify the Bragg diffractions shown in Figure S2 in the Supplementary Information. The iSALE shock physics code is not fully open-source, but is distributed on a case-by-case basis to academic users in the impact community for non-commercial use. A description of the application requirements can be found at the iSALE website (https://isale-code.github.io/terms-of-use.html). The M-ANEOS package is available from Thompson et al. (2019). The list of input parameters for the iSALE computations can be found in the Supplementary Information. The data supporting the figures in the main text are available in Ono et al. (2022).

**Acknowledgements**

We thank the developers of iSALE, including G. Collins, K. Wünnemann, B. Ivanov, J. Melosh, and D. Elbeshausen. We also thank Tom Davison for the development of pySALEPlot. We thank H. Yoshida for help with image acquisition and the miXcroscopy system (Figure S1). We also thank K. Kainuma for help with the XRD analysis (Figure S5). We appreciate Juulia-Gabrielle Moreau and an anonymous referee for their careful reviews that helped greatly improve the manuscript, and the editor for handling the manuscript. This research was supported by JSPS KAKENHI grant JP19H00726. K.K. is supported by JSPS KAKENHI grant JP18KK0092 and JP21H01140. K.K. and T.N. are supported by JSPS KAKENHI grant JP21K18660. T.M. is supported by JSPS KAKENHI grant 21K18645.

**References**

Amsden, A., Ruppel, H., & Hirt, C. (1980). SALE: A simplified ALE computer program for fluid flow at all speeds. *Los Alamos National Laboratories Report*, LA-8095:101p.

Bowling, T. J., Johnson, B. C., Wiggins, S. E., Walton, E. L., Melosh, H. J. & Sharp, T. G. (2020). Dwell time at high pressure of meteorites during impact ejection from Mars. *Icarus*, *343*, 113689.

Ivanov, B. A., Melosh, H. J., & Pierazzo E. (2010). Basin-forming impacts: reconnaissance modeling. *Geological Society of America Special Papers* 465, R.L. Gibson, W.U. Reimold (Eds.), Large Meteorite Impacts and Planetary Evolution IV, pp. 29-49"

Collins, G. S., Elbeshausen, D., Davison, T. M., Wünnemann, K., Ivanov, B. A., & Melosh, H. J. (2016). iSALE-Dellen manual. *Figshare*, https://doi.org/10.6084/m9.figshare.3473690.v2

Ivanov, B. A., de Niem, D., & Neukum, G. (1997). Implementation of dynamic strength models into 2-D hydrocodes: Applications for atmospheric breakup and impact cratering. *Int. J. Impact Eng.*, *20*, 411–430.

Ivanov, B. A., Langenhorst, F., Deutsch, A., and Hornemann, U. (2002). How strong was impact-induced $CO_2$ degassing in the Cretaceous-Tertiary event? Numerical modeling of shock recovery experiments. *in* Catastrophic event and mass extinctions: Impacts and beyond edited by C. Koeberl and K. G. MacLeod, *Geological Society of America Special Paper*, *356*, 587-594.

Ivanov, B. A., Melosh, H. J. & Pierazzo, E. (2010). Basin-forming impacts: Reconnaissance modeling. *in* Large Meteorite Impacts and Planetary Evolution IV edited by R. L. Gibson & W. U. Reimold, *Geological Society of America Special Paper*, *465*, 29-49.

Kanemaru R., Imae N., Yamaguchi A., Takenouchi A., & Nishido H. (2020). Estimation of shock degrees of eucrites using X-ray diffraction and petrographic methods. *Polar Science*, *26*:100605.

Kurosawa, K., Nagaoka, Y., Senshu, H., Wada, K., Hasegawa, S., Sugita, S., & Matsui, T. (2015). Dynamics of hypervelocity jetting during oblique impacts of spherical projectiles investigated via ultrafast imaging. *Journal Geophysical Research-Planets*, *120*, 1237–1251.

Kurosawa, K. & Genda, H. (2018). Effects of friction and plastic deformation in shock-comminuted damaged rocks on impact heating. *Geophysical Research Letters*, *45*, 620-626, https://doi.org/10.1002/2017GL076285

Kurosawa, K., Genda, H., Azuma, S., & Okazaki, K. (2021). The role of post-shock heating by plastic deformation during impact devolatilization of calcite ($CaCO_3$). *Geophysical Research Letters*, *48*, e2020GL091130.

Kurosawa, K., Moriwaki, R., Yabuta, H., Ishibashi, K., Komatsu, G., & Matsui, T. (2021). Ryugu's observed volatile loss did not arise from impact heating alone. *Communications Earth & Environment*, *2*, 146, https://doi.org/10.1038/s43247-021-00218-3

Kurosawa, K., Ono, H., Niihara, T., Sakaiya, T., Kondo, T., Tomioka, N., Mikouchi, T., Genda, H., Matsuzaki, T., Kayama, M., Koike, M., Sano, Y., Murayama, M., Satake, W., & Matsui, T. (2022). Shock recovery with decaying compressive pulses: Shock effects in calcite ($CaCO_3$) around the Hugoniot elastic limit. *Journal of Geophysical Research: Planets*, *127*, e2021JE007133. https://doi.org/10.1029/2021JE007133

Lafuente B., Downs R. T., Yang H., & Stone N. (2015). The power of databases: the RRUFF project. In: Highlights in Mineralogical Crystallography, T. Armbruster and R. M. Danisi, eds., pp 1-30.

Macke, R. J., Britt, D. T., & Consolmagno, G. J. (2011). Density, porosity, and magnetic susceptibility of achondritic meteorites. *Meteoritics & Planetary Science*, *46*, 311-326.

McSween, H. Y., Mittlefehldt, D. W., Veck, A. W., Mayne, R. G., & McCoy, T. J. (2011). HED meteorites and their relationship to the geology of Vesta and the Dawn Mission. *Space Sci. Rev.*, *163*, 141-174. DOI 10.1007/s11214-010-9637-z

Mittlefehldt, D. W. (2015). Asteroid (4) Vesta: I. The howardite-eucrite-diogenite (HED) clan of meteorites. *Chemie der Erde*, *75*, 155-183.

Miyahara M., Ohtani, E., Yamaguchi, A., Ozawa, S., Sakai, T., & Hirao, N. (2014). Discovery of coesite and stishovite in eucrite. *Proceedings of the National Academy of Sciences of the United States of America*, *111*, 10939-10942.

Nakazawa, S., Watanabe, S., Iijima, Y., & Kato, M. (2002). Experimental investigation of shock wave attenuation in basalt. *Icarus*, *156*, 539–550.

Ono, H., Kurosawa, K., Niihara, T., Mikouchi, T., Tomioka, N., Isa J., Kagi, H., Matsuzaki, T., Sakuma, H., Genda, H., Sakaiya, T., Kondo, T., Kayama, M., Koike, M., Sano, Y., Murayama, M., Satake, W., & Matsui, T.. (2022). Shock recovery with decaying compressive pulses: Shock melt veins in basaltic rocks [Data set]. Zenodo. https://doi.org/10.5281/zenodo.7022698

Pierazzo, E., Artemieva, N. A., Ivanov, B. A. (2005). Starting conditions for hydrothermal systems underneath Martian craters: Hydrocode modeling. *in Large meteorite impacts III*, eds., Kenkmann, T., Hörz, F., and Deutsch, A.,: *Geological Society of America Special Paper, 384*, p. 443–457.

Quintana, S. N., D. A. Crawford, and P. H. Schultz (2015), Analysis of impact melt and vapor production in CTH for planetary applications, *Procedia Engineering*, *103*, 499-506.

Rubin, A. E. (2015). Maskelynite in asteroidal, lunar and planetary basaltic meteorites: An indicator of shock pressure during impact ejection from their parent bodies. *Icarus*, *257*, 221-229.

Sato M., Kurosawa K., Kato S., Ushioda M., and Hasegawa S. (2021). Shock Remanent Magnetization Intensity and Stability Distributions of Single-Domain Titanomagnetite-Bearing Basalt Sample Under the Pressure Range of 0.1–10 GPa. *Geophysical Research Letters*, *48*, 8.

Schmedemann, N. Kneissl, T., Ivanov, B. A., Michel, G. G., Wagner, R. J., Neukum, G., O. Ruesch, Hiesinger, H., Krohn, K., Roatsch, T., Preusker, F., Sierks, H., Jaumann, R., Reddy, V., Nathues, A., Walter, S. H. G., Neesemann, A., Raymond, C. A. & Russel, C. T. (2014). The cratering record, chronology, and surface age of (4) Vesta in comparison to smaller asteroids and the ages of HED meteorites. *Planetary and Space Science*, *103*, 104-130.

Stöffler, D., Keil, K., & Scott, E. R. (1991). Shock metamorphism of ordinary chondrites. *Geochimica et Cosmochimica Acta*, *55*, 3845–3867.

Stöffler, D., Hamann, C., & Metzler, K. (2018). Shock metamorphism of planetary silicate rocks and sediments: Proposal for an updated classification system. *Meteoritics and Planetary Science*, *53*, 5-49.

Sugita, S. & Schultz, P. H. (2003). Interactions between impact-induced vapor clouds and the ambient atmosphere: 2. Theoretical modeling. *Journal of Geophyical Research*, *108*, 5052.

Thompson, S. L., & Lauson, H. S. (1972). Improvements in the Chart-D radiation hydrodynamic code III: Revised analytical equation of state. pp. *SC-RR-71 0714* 119 pp., Sandia Laboratories, Albuquerque, NM.

Thompson, S. L., Lauson, H. S., Melosh, H. J., Collins, G. S., & Stewart, S. T. (2019, November 1). M-ANEOS (Version 1.0). Zenodo. https://doi.org/10.5281/zenodo.3525030

Tillotson, J. H. (1962). Metallic equations of state for hypervelocity impact. Technical Report *GA-3216*, General Atomic Report.

Wünnemann, K., Collins, G. S., & Melosh, H. J. (2006). A strain-based porosity model for use in hydrocode simulations of impacts and implications for transient crater growth in porous targets. *Icarus*, *180*, 514–527.

Supporting Information for

**Experimentally shock-induced melt veins in basalt: Improving the shock classification of eucrites**

**Contents of this file**
    Texts S1 to S10
    Figures S1 to S12
    Tables S1 to S6
    Captions for Movies S1 to S3

**Additional Supporting Information (files uploaded separately)**
    Movies S1 to S3

**Text S1. Sample descriptions**

    The basalt was collected from Inner Mongolia (China) from the same location as the basalt used by Sato et al. (2021). Figure S1 shows transmitted light images of the thin-section of an unshocked sample obtained with an optical microscope (Utokyo and JEOL; miXcroscopy). The basalt sample consists mainly of fine-grained (80–200 μm in size) plagioclase, pyroxene, and olivine in the matrix, and coarse-grained pyroxene and olivine phenocrysts, and opaque minerals. The basalt contains titanomagnetite (0.7 wt.%), which is a magnetic mineral [Sato et al., 2021]. As mentioned in the main text, the density of this basalt sample is 2.94 Mg m$^{-3}$, which is comparable to that of known eucrites (2.66 – 3.12 Mg m$^{-3}$) [Macke et al., 2011]. We conducted macroscopic and microscopic observations on the unshocked basalt to investigate the nature of porosity in the sample. It is widely known that both microscopic and macroscopic porosities in shocked media effect the momentum and energy partitioning between the projectile and target [e.g., Ahrens & O'Keefe, 1972; Wünnemann et al., 2006, 2008; Davison et al., 2010, 2014, 2016; Güldemeister et al., 2013, 2022; Kowitz et al., 2013; Bland et al., 2014; Moreau et al., 2019]. The hand specimens contain nearly spherical voids on a millimeter scale. The voids are rare and occur at 1–10 per 1000 cm$^2$ of sample area. The sample is closely packed, and fractures were not evident during macroscopic observation. As such, the rare vesicles were visible with the naked eye or by microscopic observations.

    We conducted a microscopic observation of the unshocked sample with a scanning electron microprobe (SEM; JEOL JSM-6510LA; Chiba Institute of Technology) to investigate the nature of porosity in the sample. The accelerating voltage was set to 15 kV. Figure S2 shows back-scattered electron images of the unshocked sample. The sample is tightly packed and open pores and cracks are absent along mineral boundaries down to the sub-micron scale although some fractures occur within minerals with a certain number density. In such a tightly packed rock, it should be possible to locate macroscopic vesicles if they existed. We note that such voids were not seen during petrographic observations after the experiments. Consequently, we neglected the effects of porosity compaction on the impact outcomes of this study.

    We investigated the mineral compositions and modal mineralogy of the sample with the same electron probe microanalyzer as used for the chemical analysis described in the main

text. Figures S3–S5 show the feldspar and pyroxene compositions, respectively, for both our sample and eucrites. The chemical compositions of the plagioclase and pyroxene are $An_{54.4–65.7}Ab_{33.0–44.3}$ and $Fs_{14.0–33.0}En_{35.2–49.3}Wo_{30.9–39.4}$, respectively. The modal mineralogy was estimated from X-ray elemental maps of the unshocked basalt. The mineral compositions of silicates in our sample are summarized in Table S1. There are three differences between our sample and eucrites. Firstly, the Na/Ca ratio of the plagioclase in our sample is greater than that of eucrites. Secondly, the ratio of felsic to mafic minerals in our sample is greater than that of basaltic eucrites. Thirdly, the fraction of opaque minerals in our sample is slightly higher than that of eucrites. Despite these differences, we chose the Inner Mongolian basalt because its density is comparable to that of eucrites and it is homogeneous and nearly non-vesicular.

We also conducted X-ray diffraction (XRD; SmartLab, Rigaku Corporation) analyses at the Chiba Institute of Technology, Narashino, Japan. The XRD measurements were undertaken with Cu $K\alpha$ radiation with a $\lambda = 0.154$ nm at 45 kV and 200 mA, and a scan rate of $2°$ min$^{-1}$. The indices of the Bragg peaks were assigned with PDXL software (Rigaku). The software includes the International Center for Diffraction Data (ICCD) PDF-2 database, and the inorganic crystal structure database provided by the Crystallographic Society of Japan. Figure S6 shows the XRD pattern of the unshocked basalt. We identified the Bragg reflections of Ti-bearing Fe oxides (ilmenite) along with those of the silicates.

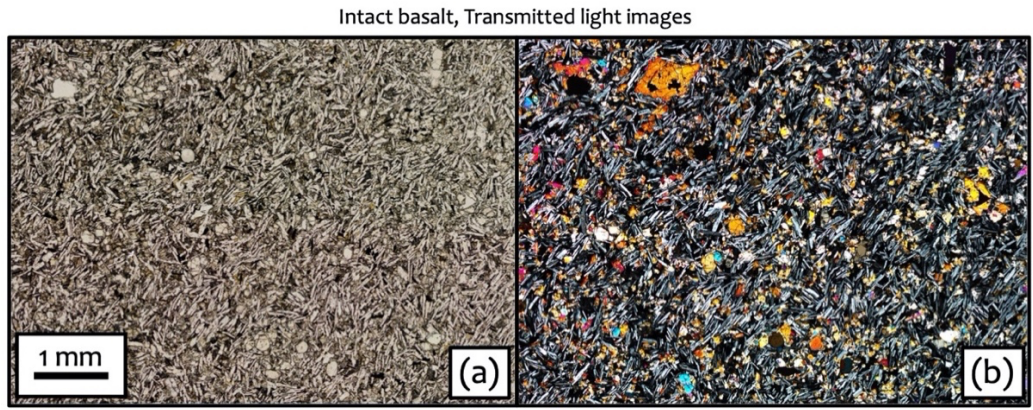

**Figure S1.** Transmitted light images of the unshocked basalt sample. (a) Plane- and (b) crossed- polarized light.

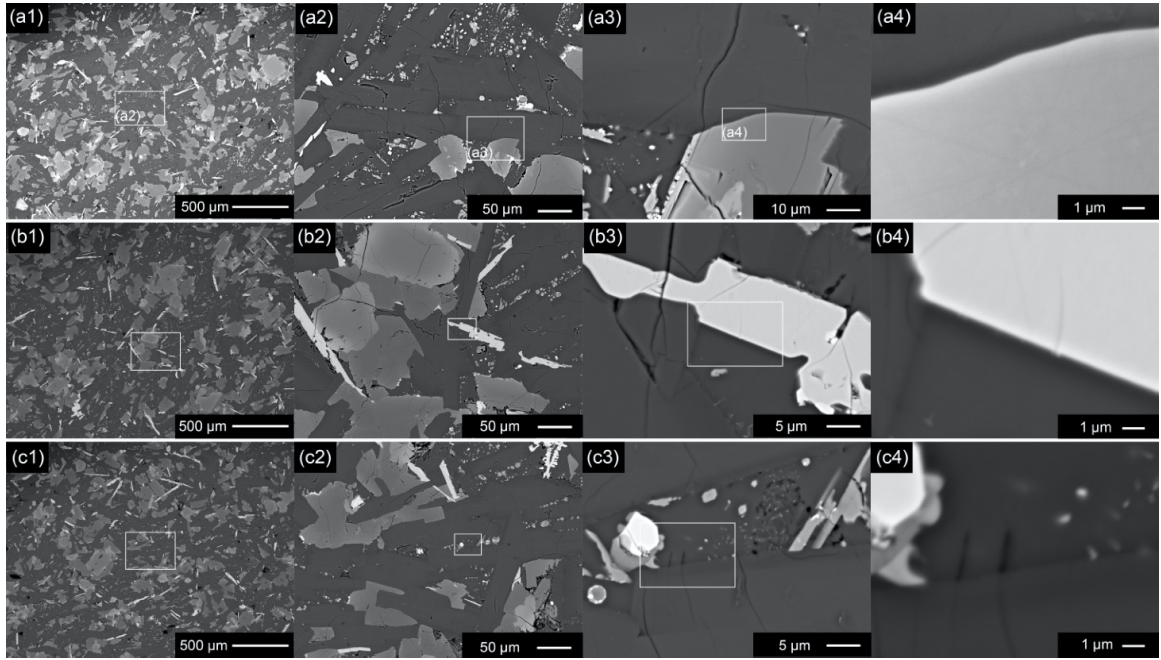

**Figure S2**: Back-scattered electron (BSE) images of the basalt that were used as the starting materials for the experiments. Three locations were randomly selected from a thick section of the starting material. The locations are labeled a, b, and c and shown in the top, middle, and bottom panels. We arranged the images by order of magnification from left (1) to right (4). The white boxes in the lower resolution images (1-3) correspond to the magnified images their right.

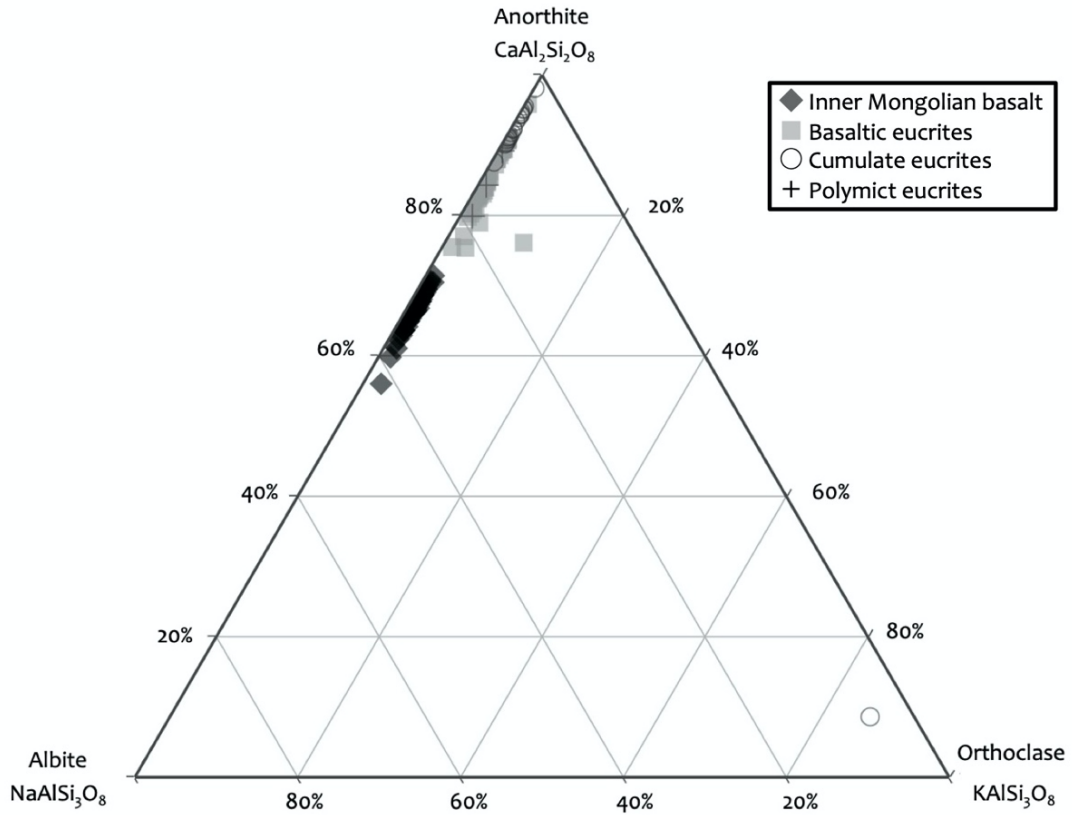

**Figure S3.** Feldspar chemical compositions. The measured feldspar compositions of the Unshocked basalt sample obtained with an electron probe microanalyzer are plotted, as well as data for basaltic, cumulate, and polymict eucrites. The eucrite data are from Duke and Silver (1967), Takeda et al. (1976, 1983), Warren et al. (1990), Mayne et al. (2009), Barrat et al. (2011), Mittlefehldt (2015, 2022) and references therein, Yamaguchi et al. (2017), and Ono et al. (2019).

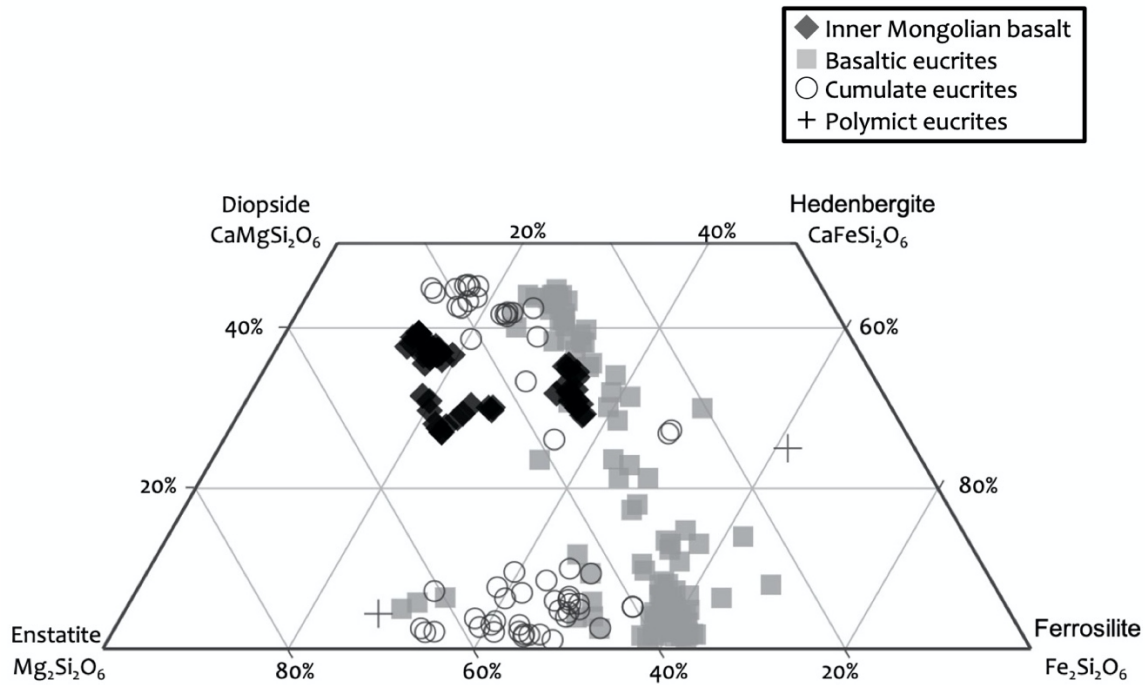

**Figure S4.** Same as Figure S2, except for pyroxene.

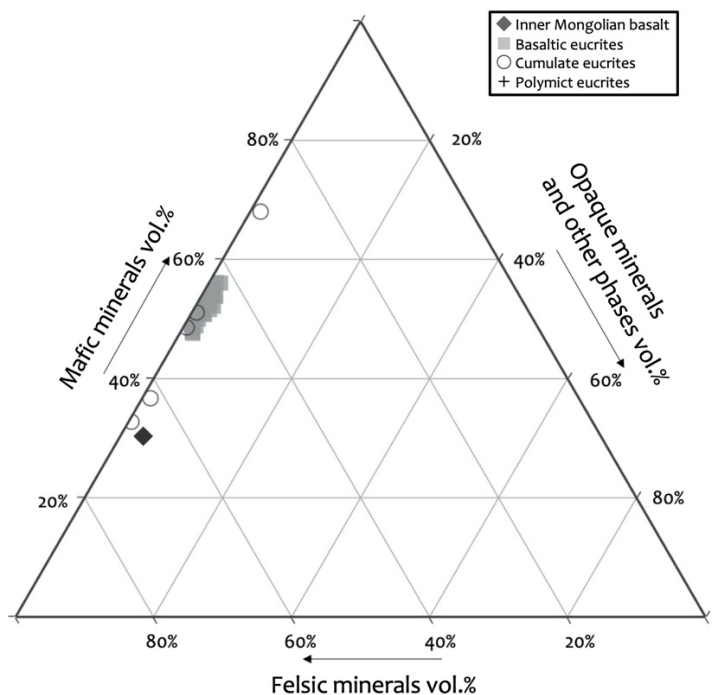

**Figure S5.** Modal mineralogy. Modal mineralogy of the Inner Mongolian basalt estimated from X-ray elemental maps obtained with an electron probe microanalyzer. The mafic minerals, felsic minerals, and opaque and other phases include olivine and pyroxene, Al- and/or Si-rich minerals, and ilmenite, chromite, troilite, metals, and other minor phases, respectively. The eucrite data are from the references listed in the caption to Figure S2.

**Table S1.** Chemical compositions of the minerals in the unshocked basalt measured by electron probe microanalyzer. Abbreviations: Pl = plagioclase, HC-Px = high-Ca pyroxene, LC-Px = low-Ca pyroxene, Ol = olivine, n.d. = not detected, D.L. (wt.%) = detection limit, Fs = ferrosilite, En = enstatite, Wo = wollastonite, An = anorthite, Ab = albite, Or = orthoclase, Fa = fayalite, and Fo = forsterite.

| (wt.%) | Pl | LC-Px | HC-Px | Ol | D.L. (wt.%) |
|---|---|---|---|---|---|
| $SiO_2$ | 52.64 | 51.89 | 52.3 | 37.69 | 0.04 |
| $Al_2O_3$ | 28.95 | 1.55 | 2.21 | n.d. | 0.05 |
| $TiO_2$ | 0.09 | 0.96 | 0.90 | n.d. | 0.05 |
| FeO | 0.57 | 14.33 | 8.82 | 27.09 | 0.03 |
| MnO | n.d. | 0.35 | 0.16 | 0.32 | 0.03 |
| MgO | 0.15 | 17.46 | 15.94 | 34.80 | 0.03 |
| CaO | 12.5 | 13.11 | 18.87 | 0.31 | 0.02 |
| $Na_2O$ | 4.19 | 0.19 | 0.30 | n.d. | 0.06 |
| $K_2O$ | 0.17 | n.d. | n.d. | n.d. | 0.03 |
| $Cr_2O_3$ | n.d. | 0.10 | 0.58 | n.d. | 0.05 |
| Total | 99.26 | 99.94 | 100.08 | 100.21 | |
| | | | | | |
| *Fs* | | 23.0 | 14.4 | | |
| *En* | | 50.0 | 46.3 | | |
| *Wo* | | 27.0 | 39.4 | | |
| | | | | | |
| *An* | 61.6 | | | | |
| *Ab* | 37.4 | | | | |
| *Or* | 1.0 | | | | |
| | | | | | |
| *Fa* | | | | 30.3 | |
| *Fo* | | | | 69.3 | |

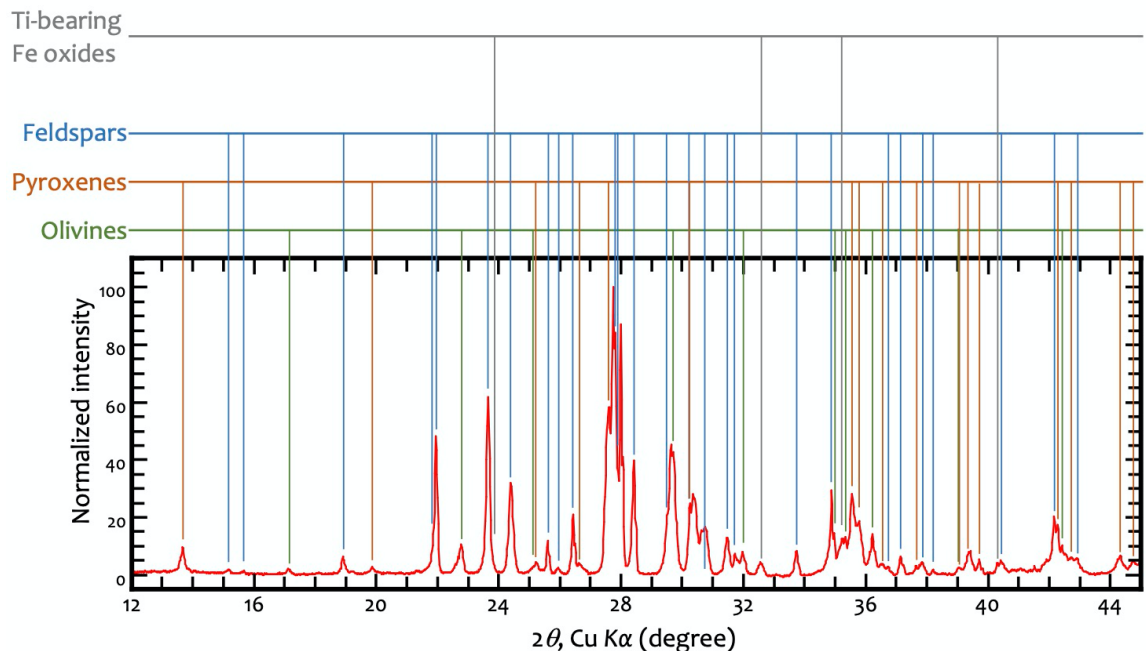

**Figure S6.** X-ray diffraction pattern of the powdered unshocked basalt. The detected peaks are indicated on the figure.

**Text S2. Analytical conditions during the microscopic observations**

We used a scanning electron microscope (SEM) and a field emission scanning electron microscope (FE-SEM). The accelerating voltage and working distance of the SEMs were 15 kV and 10 mm, respectively. We also used two electron probe microanalyzers. The analytical conditions were: accelerating voltage = 15 kV, beam current = 12 nA, and beam diameter = 2 μm. Electron back-scattered diffraction (EBSD) analysis was conducted with a 15 kV accelerating voltage, 25 mm working distance, and sample holder tilting by 70°.

The Raman spectra were obtained with a micro-Raman spectrometer comprising a confocal optical microscope (Olympus, BX51) with a 100× objective lens, 514.5 nm Ar ion laser (Melles Griot 543-AP-A01), single polychromator (Bruker Optics 500 Imaging Spectrograph) equipped with 1200 lines mm$^{-1}$ grating, and Peltier-cooled CCD detector (Andor Technology DU401A-BG-DD) at the Geochemical Research Center of the University of Tokyo, Japan. The laser power was set to 7.8 mW on the polished thin-section, and the acquisition time was 150 s (10 s × 15 repeats). The laser spot on the thin-section is oval in shape, with a minor axis of 1 μm and major axis of 5 μm.

**Text S3. Shock physics modeling**

We used the iSALE-Dellen package [Collins et al., 2016] to estimate the peak pressure and peak temperature distributions as a function of distance from the epicenter, as described in the main text (section 2.3). Two-dimensional cylindrical coordinates were used to model the vertical impacts performed in the impact experiment. The grid spacing was set to 50 μm. The corresponding numbers of cells per projectile radius (CPPR) were 46 and 20 for the polycarbonate and Ti projectiles, respectively. The CPPR value is equivalent to the spatial resolution, and CPPR values of 46 and 20 were sufficiently high to accurately estimate the peak pressure distribution [Pierazzo et al., 2008; Kurosawa et al., 2022]. We used the Tillotson EOS [Tillotson, 1962] for polycarbonate [Sugita and Schultz, 2003] and Ti [Tillotson, 1962], and ANEOS [Thompson and Lauson, 1972] for basalt [Pierazzo et al., 2005; Sato et al., 2021]. The rock strength and von-Mises strength models in the iSALE package (Collins et al., 2016) were used for the basalt and Ti container, respectively. We inserted Lagrangian tracer particles into each computational cell. The maximum recorded peak pressure and peak temperatures experienced by each individual tracer were stored during the simulation. The input parameters of the material models and calculation settings are summarized in Tables S2–S3. Note that the Ti container in the numerical simulation was set to be free from the computational boundaries. As such, the selection of the boundary conditions does not affect the results.

**Table S2.** Input parameters for the material models. Note that the parameter sets pertaining to the basalt target and Ti container used here are the same as those used by Bowling et al. (2020) and Kurosawa et al. (2022).

|  | Projectile | Target | Container |
|---|---|---|---|
| EOS type | Tillotson | ANEOS | Tillotson |
| Material | Polycarbonate or Ti | Basalt | Ti |
| Strength model | HYDRO | Rock | Von Mises |
| Poisson's ratio | 0.5 | 0.25 | 0.34 |
| Melting temperature (K) | Not used. | 1360 | 1941 |
| Thermal softening parameter | Not used. | 0.7 | 1.2 |
| Simon parameter, A (GPa) | Not used. | 1.8 | 7 |
| Simon parameter, C | Not used. | 7.3 | 7.19 |
| Cohesion (undamaged) (MPa), $Y_{coh,i}$ | Not used. | 20 | 500 |
| Cohesion (damaged) (kPa), $Y_{coh}$ | Not used. | 10 | 10 |
| Internal friction (undamaged), $\mu_{int}$ | Not used. | 1.4 | Not used. |
| Internal friction (damaged), $\mu_{dam}$ | Not used. | 0.6 | Not used. |
| Limiting strength (GPa), $Y_{limit}$ | Not used. | 2.5 | Not used. |
| Minimum failure strain | Not used. | $10^{-4}$ | Not used. |
| Constant for the damage model | Not used. | $10^{-11}$ | Not used. |
| Threshold pressure for the damage model (MPa) | Not used. | 300 | Not used. |

**Table S3.** Numerical model settings. These are the same as used by Kurosawa et al. (2022), except for the CPPR value for shot #492 and the impact velocities.

| | |
|---|---|
| Computational geometry | Cylindrical coordinates |
| Number of computational cells in the $R$ direction | 700 |
| Number of computational cells in the $Z$ direction | 1200 |
| Cells per projectile radius (CPPR) | 46 (for shots #471 and #476) |
| | 20 (for shot #492) |
| Grid spacing (m grid$^{-1}$) | $5 \times 10^{-5}$ |
| Artificial viscosity, $a_1$ | 0.24 |
| Artificial viscosity, $a_2$ | 1.2 |
| Impact velocity (km s$^{-1}$) | 7.31 (for shot #471) |
| | 7.01 (for shot #476) |
| | 6.92 (for shot #492) |
| Low-density cutoff (kg m$^{-3}$) | 1 |

**Text S4. Validity of the shock physics modeling**

The peak pressure distributions in the shocked samples were only estimated with the iSALE package. We conducted a validity test of the iSALE results by comparing these with the published experimental data of Nakazawa et al. (2002), who measured the attenuation rate of the peak pressures in basalt blocks with gauges. We modeled their "16 GPa experiment" with iSALE. In this experiment, three Cu cylinders with a diameter of 9 mm and length of 9 mm were accelerated to 1.6–1.7 km s$^{-1}$. Those authors measured the peak stress $\sigma_{ZZ,peak}$ normal to the gauges and parallel to the impact direction at two different initial distances from the impact points, and conducted a total of three shots. The pressures at the impact points in the three shots are mostly identical to 16 GPa. We used the same EOS and material models pertaining to the basalt target, except that the reference density was adjusted for their sample (2.7 Mg m$^{-3}$). The Tillotson EOS with the parameters for Cu [Tillotson, 1962] and the Johnson–Cook strength model with the parameter set for OFHC Cu [Johnson and Cook, 1983] were used for the Cu cylinder. The polycarbonate sabot was also included in our numerical model as a perfect fluid. The Tillotson EOS for polycarbonate [Sugita and Schultz, 2003] was used. We inserted Lagrangian tracer particles into each computational cell. We stored the temporal pressures $P$ and values of the deviatoric stress tensor in the $Z$ direction $s_{zz}$ on the tracers. The stress $\sigma_{ZZ}$ can be calculated as $\sigma_{ZZ} = P + s_{zz}$ [e.g., Collins et al., 2013, pp. 256]. The input parameters and calculation settings are summarized in Tables S4–S5. The computational grids were sufficiently large that we could ignore the effects of wave refraction from the boundaries. Therefore, the selection of the boundary conditions does not affect the results.

Figure S7 shows the initial density and distribution of $\sigma_{ZZ,peak}$ as a function of the initial locations of the Lagrangian tracers (i.e., provenance plots). The gauge locations in the "16 GPa experiment" are also shown. Figure S8 shows $\sigma_{ZZ,peak}$ as a function of the initial distance from the impact point along with the experimental data from Nakazawa et al. (2002). For reference, we also show the peak pressure $P_{peak}$ distribution in the figure. We plotted $\sigma_{ZZ,peak}$ and $P_{peak}$ on the tracers initially located 5–10 cells from the symmetry axis. We found that the calculated $\sigma_{ZZ,peak}$ from iSALE reproduces the experimental data at stresses of >10 GPa. In contrast, the accuracy of the calculated $\sigma_{ZZ,peak}$ values tends to decrease with increasing distance. At the farthest gauge, the calculated $\sigma_{ZZ,peak}$ is 2–3 times higher than the experimental data. Such an overestimation far from the impact point also occurred in the experiment of Kurosawa et al. (2022). The ANEOS basalt used in this study [Sato et al., 2021] does not include the effects of the sudden decrease in the wave propagation speed at the Hugoniot elastic limit, as with the ANEOS calcite [Pierazzo et al., 1998]. This simplification in the EOS model leads to the omission of the wave splitting that is frequently observed in actual rocky materials [e.g., Melosh, 1989, pp. 36]. Wave splitting was not reproduced in the iSALE simulation with the ANEOS basalt, resulting in momentum transport with a longer dwell time, such that the calculated $\sigma_{ZZ,peak}$ is an overestimation at the gauges located farther than from 10 mm from the epicenter at stresses of below 3 GPa. Although the peak pressures shown in Fig. 1a in the main text would be an overestimation, particularly at the far end of the shocked basalt target, the peak pressure distributions around the epicenter and at >10 GPa (Fig. 1b in the main text) are reliable.

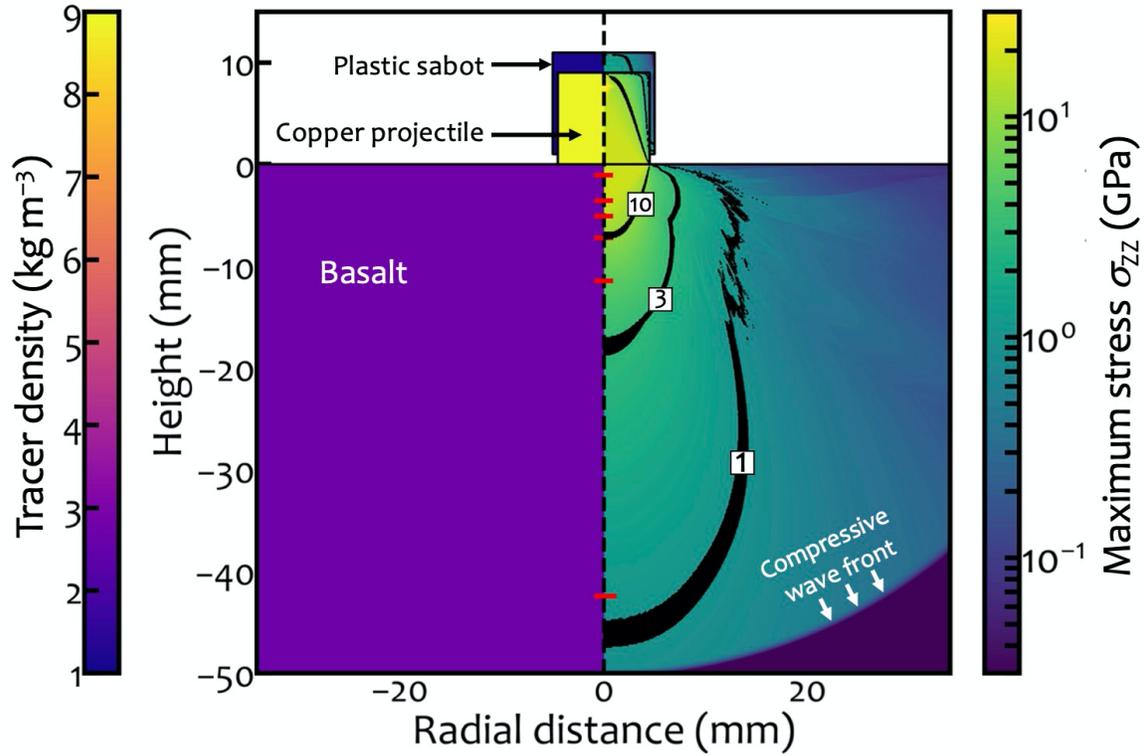

**Figure S7.** Provenance plots of the initial density (left) and maximum stress (right). The locations of the stress gauges are shown as red lines.

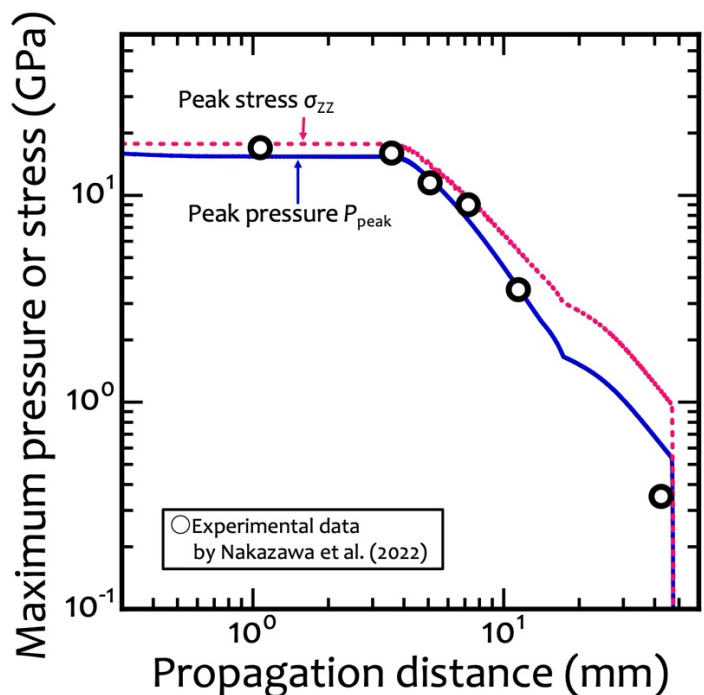

**Figure S8.** Maximum pressure (blue solid line) and stress (pink dashed line) as a function of propagation distance. The experimental data of Nakazawa et al. (2002) are shown as black open circles.

**Table S4.** Input parameters of the material models for the Cu cylinder and polycarbonate sabot.

|  | Projectile | Sabot |
|---|---|---|
| EOS type | Tillotson | Tillotson |
| Material | Cu | Polycarbonate |
| Poisson's ratio | 0.34[a] | 0.5 |
| Melting temperature (K) | 1356[b] | Not used. |
| Simon parameter, $a$ (GPa) | 49.228[b] | Not used. |
| Simon parameter, $c$ | 1.027[b] | Not used. |
| Specific heat (J K$^{-1}$ kg$^{-1}$) | 392.5[c] | Not used. |
| Johnson–Cook parameter, $A$ (MPa) | 90[d] | Not used. |
| Johnson–Cook parameter, $B$ (MPa) | 292[d] | Not used. |
| Johnson–Cook parameter, $N$ | 0.31[d] | Not used. |
| Johnson–Cook parameter, $C$ | 0.025[d] | Not used. |
| Johnson–Cook parameter, $m$ | 1.09[d] | Not used. |
| Reference temperature (K) | 293 | Not used. |

a. Köster and Franz (1961).
b. The pressure-dependent melting temperature of Cu from Japel et al. (2005) was fitted using the Simon equation (e.g., Poirier, 1991; Wünnemann et al., 2008).
c. The Dulong–Petit values were used.
d. Johnson and Cook (1983).

**Table S5.** Numerical model settings used in the validity test. The details of the variables can be found in Collins et al. (2016).

| | |
|---|---|
| Computational geometry | Cylindrical coordinates |
| Number of computational cells in the $R$ direction | 1200 |
| Number of computational cells in the $Z$ direction | 3000 |
| Number of extension cells in the $R$ direction | 100 |
| Number of extension cells in the $Z$ direction | 140 |
| Extension factor | 1.02 |
| Largest size of the extension cell | 20 times that of the normal cell |
| Cells per projectile radius (CPPR)[b] | 100 (Cu projectile) |
| | 111 (polycarbonate sabot) |
| Grid spacing (m grid$^{-1}$) | $4.5 \times 10^{-5}$ |
| Impact velocity (km s$^{-1}$) | 1.65 |
| Layer position | 2800 cells from the bottom of the computational domain |
| Artificial viscosity, $a_1$ | 0.24 |
| Artificial viscosity, $a_2$ | 1.2 |

**Text S5. Three-dimensional X-ray computational tomography (3D-XCT) of a shocked basalt cylinder**

We observed the three-dimensional damage structure of the recovered sample from shot #476 with a micro-focus X-ray computational tomography (XCT) instrument (Zeiss Xradia 410 Versa) at the Center for Advanced Marine Core Research, Kochi University, Japan. The imaging conditions were the same as those used by Kurosawa et al. (2022). An X-ray tube with a W target was employed. The spatial resolution was limited to ~60 μm. Collecting optics with a 0.4× objective were used. The X-ray tube voltage and power were 140 kV and 8.0 W, respectively. The exposure time and number of projections were 2 s and 1601, respectively. Figure S9 shows four 3D XCT images of cross-sections of the shocked sample with different azimuth angles. We observed open cracks with widths of >100 μm, and we confirmed that there is no clear dependence of the azimuth angle on the damage structures. Multiple-concentric open cracks were clearly observed beneath the floor of the impact-generated depression. Relatively large-scale radial failures were not developed in our sample in contrast to Winkler et al. (2018), which studied sub-surface damage structure in shocked quartzite sample. Although we only analyzed a single specific cross-section of the shocked samples, as described in the main text, the selection of the specific azimuth angle from the shocked sample does not change the main conclusion of this study.

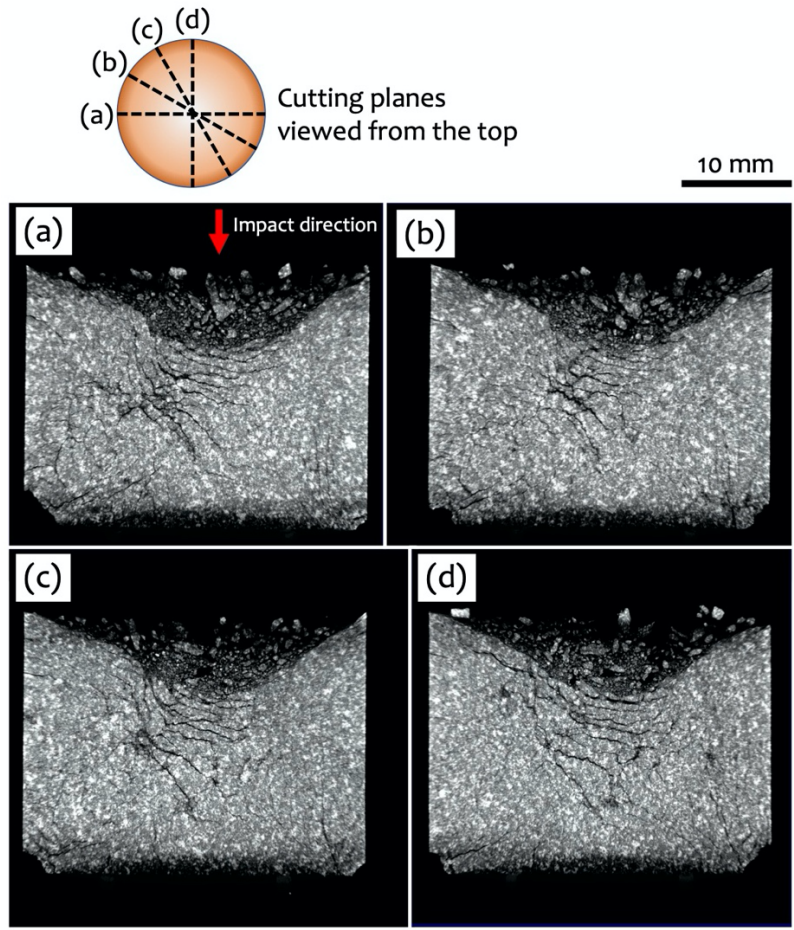

**Figure S9.** Four images obtained by three-dimensional X-ray computational tomography at different azimuth angles. The cutting plane of each panel from (a) to (d) is schematically shown at the top of the figure.

**Text S6. Microscopic observations of polished thin-sections**

We present a few examples of the observations under a polarizing microscope as Supplementary Movies S1–S3. We observed the changes in the extinction patterns of the silicates as the stage on the polarizing microscope was rotated. Although we only show movies taken of a polished thin-section from shot #471, the extinction behavior is similar for shot #492.

**Captions for the Supplementary Movies**
**Movie S1.** Transmitted light image in cross-polarized light taken near the epicenter. The field of view (FOV) corresponds to the white box (i) in Figure S10. The size of the image is ~1.3 × 0.7 mm.

**Movie S2.** Same as Movie S1, except that the FOV corresponds to the white box (ii) in Figure S10.

**Movie S3.** Same as Movie S1, except that a region on the unshocked basalt sample is shown.

**Text S7. Entire map of the shocked sample from shot #471**

We made a mosaic with a SEM of the shocked sample from shot# 471 (Figure S10). Although we analyzed mainly the region around the epicenter (i.e., the red rectangle in the figure), open cracks were produced during the experiment as observed in the 3D-XCT images (Text S5).

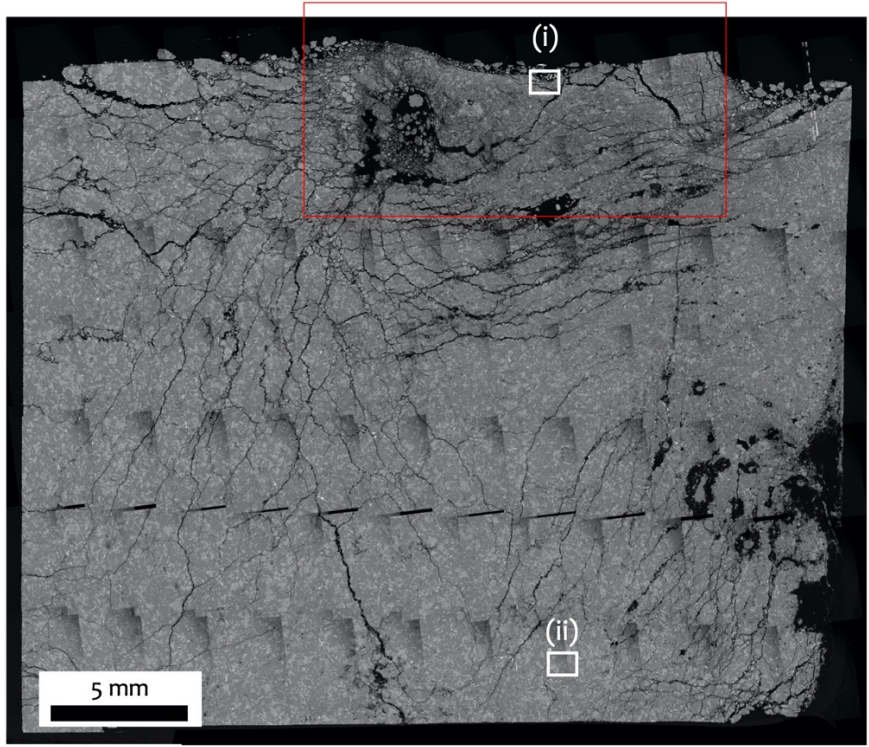

**Figure S10.** Back-scattered electron image of the recovered sample from shot #471. The red rectangle indicates the region shown in Figure 1b in the main text. The white boxes (i) and (ii) are the field of views of Supplementary Movies S1 and S2, respectively.

**Text S8. An elemental map around a shock melt vein**

We obtained an elemental map around one of the shock melt veins (SMVs) from shot #471 with an electron probe microanalyzer. Figure S10a shows a back-scattered electron (BSE) image of the region outlined by the cyan box (ii) in Figure 1b in the main text, and Figure S10b shows an enlarged BSE image of the white rectangle in (a). The SMV is located at the center of (b). Figure S11c–h shows elemental maps of Ca, Mg, Fe, Al, Si, and Na, respectively, for the same region as shown in (b). Although the materials in the SMVs are mainly plagioclase, a mafic component also intrudes the SMVs (see [d–e]). In addition, Ca-rich plagioclase, which has a relatively high melting point (anorthite = 1825 K at atmospheric pressure; [Ahrens and O'Keefe, 1972]), was also melted. However, the Na concentration of the region adjacent to the SMV is still high (see [h]), indicating that Na diffusion into the melt phase did not occur. As such, the duration of the high temperature would have been rather short, although an investigation of the detailed thermal history of the SMVs is beyond the scope of this study.

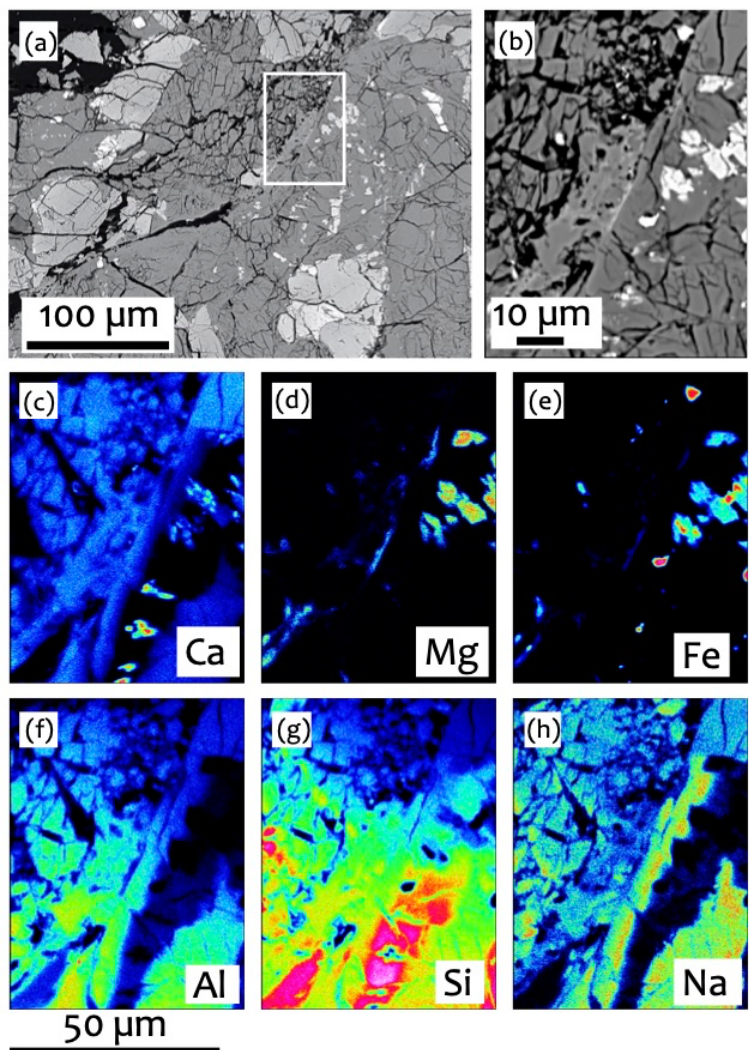

**Figure S11.** (a) A BSE image of the region shown in the blue box (ii) in Figure 1b in the main text. (b) An enlarged BSE image of the region corresponding to the white rectangle in (a). (c–h) Elemental maps of the region shown in (b).

**Text S9. Reproducibility of the experiment**

We made two polished thin-sections of the recovered samples (shots #471 and #492), although we mainly discuss the results of shot #471 in the main text. In this section, we present BSE images of the thin-section from shot #492. Figure S12 shows the region around the epicenter along with the estimated peak pressure. We also identified SMVs in the region that experienced a peak pressure of >10 GPa, similar to shot #471.

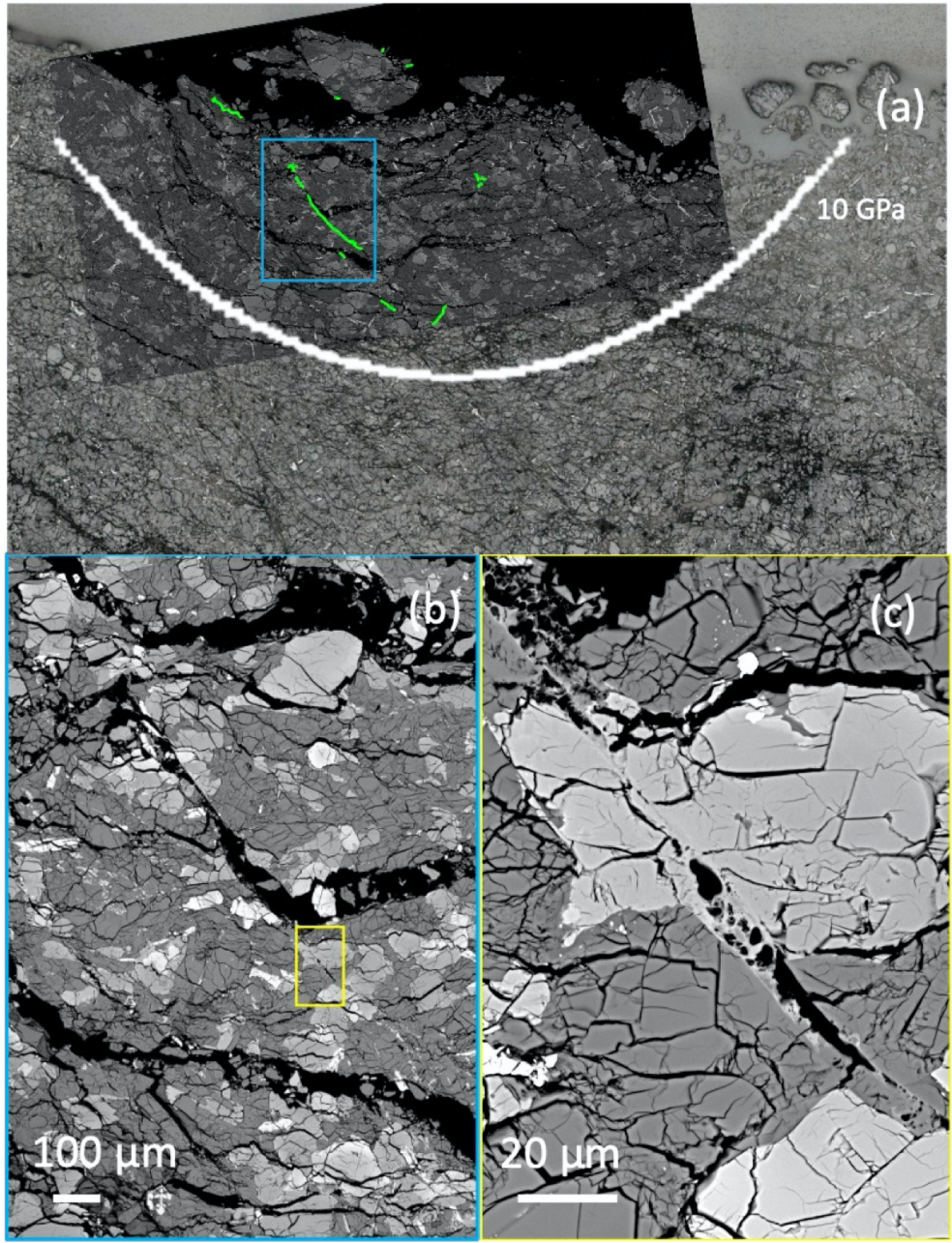

**Figure S12.** Same as Figure 1b–d in the main text, except that the thin-section from shot #492 is shown. SMVs are indicated by green lines on the BSE image superimposed on the reflected light image (a). (b) Enlarged view of the rectangular region in (a), showing SMV texture. (c) Enlarged view of the rectangular region in (b).

**Text S10. Shocked mass during impacts on Vesta**

We conducted a series of iSALE simulations to estimate the shocked masses $M_C$ and $M_{D\&E}$ classified into the shock degree C and shock degrees D–E as discussed in section 4.3 in the main text. We assumed there was a vertical impact of a basaltic projectile onto a basaltic crust with zero porosity. We employed cylindrical coordinates. The projectile was divided into 50 cells per projectile radius (CPPR). The number of CPPR is high enough to accurately estimate the peak pressure distribution [e.g., Pierazzo et al., 2008]. In the simulations, we ignored gravity because we focused only on the peak pressure distribution, which is determined during the early stage of the impact. The material model pertaining to both the projectile and target is the same as listed in the third column of Table S2 (i.e., for basalt). Given that we did not include scale-dependent physics in this series of simulations, we normalized the length and shocked mass to the projectile radius and mass. The calculation settings are summarized in Table S6.

**Table S6.** Numerical model settings used in the Vesta simulations. Note that extension zones were not used.

| | |
|---|---|
| Computational geometry | Cylindrical coordinates |
| Number of computational cells in the $R$ direction | 1000 |
| Number of computational cells in the $Z$ direction | 1200 |
| Cells per projectile radius (CPPR)[b] | 50 |
| Impact velocity (km s$^{-1}$) | 2–14 with 0.5 steps |
| Layer position | 1000 cells from the bottom of the computational domain |
| Artificial viscosity, $a_1$ | 0.24 |
| Artificial viscosity, $a_2$ | 1.2 |

It should be noted here that the known eucrites have a porosity range of 0%–20% [Macke et al., 2011]. Nevertheless, we neglected the effect of porosity on the impact outcomes for the following reasons. The origin of porosity in the known eucrites is poorly understood. The known eucrites have at least experienced shear stress during their ejection. The longitudinal stress $s_L$ during the ejection can be roughly estimated as follows:

$$s_L \sim -\rho C_L u_p \sim -\rho C_L \frac{v_{esc}}{2}, \quad (S1)$$

where $\rho \sim 2.8$ Mg m$^{-3}$ is the density of the basaltic crust, $C_L = 4.6$ km s$^{-1}$ is the longitudinal speed of sound in basalt [Sekine et al., 2008], and $v_{esc} = 0.39$ km s$^{-1}$ is the escape velocity from Vesta. In the above estimation, we assumed that the shocked materials were ejected at twice the value of $u_p$ by applying the velocity-doubling rule at the free surface. The maximum shear stress $t$ was calculated as follows [e.g., Melosh, 1989, pp. 34]:

$$\tau = -\frac{(1-2\nu)}{2(1-\nu)}\sigma_L \sim 1 \text{ GPa}, \quad (S2)$$

where $\nu = 0.25$ is the Poisson's ratio of basalt [e.g., Bowling et al., 2020]. The estimated $\tau$ is much higher than the tensile strength of basalt (10–20 MPa; e.g., Nakamura et al., 2015), implying that the eucrites would be damaged during escape from Vesta. As such, the porosity in eucrites could have been produced during impact ejection.

On a global scale, geodesic measurements of Vesta by the Dawn mission also indicated that the crust of Vesta has an average porosity of ~10% [Park et al., 2014].

However, the nature of the porosity on the millimeter to meter scale, which is comparable to the size of eucrites, on Vesta is poorly constrained. Consequently, we assumed a zero porosity for the Vestan crust, as this is the simplest assumption.

**References**

Ahrens, T. J. & O'keefe, J. D. (1972). Shock melting and vaporization of lunar rocks and minerals. *The Moon*, *4*, 214-249.

Barrat, J. A., Yamaguchi, A., Bunch, T. E., Bohn, M., Bollinger, C., & Ceuleneer, G. (2011). Possible fluid–rock interactions on differentiated asteroids recorded in eucritic meteorites. *Geochimica et Cosmochimica Acta*, *75*, 3839-3852.

Bland, P. A., Collins, G. S., Davison, T. M., Abreu, N. M., Ciesla, F. J., Muxworthy, A. R., & Moore, J. (2014). Pressure-temperature evolution of primordial solar system solids during impact-induced compaction. *Nature Communications*, 5:5451.

Bowling, T. J., Johnson, B. C., Wiggins, S. E., Walton, E. L., Melosh, H. J. & Sharp, T. G. (2020). Dwell time at high pressure of meteorites during impact ejection from Mars. *Icarus*, *343*, 113689.

Collins, G. S., Elbeshausen, D., Davison, T. M., Wünnemann, K., Ivanov, B. A., & H. J., Melosh (2016). iSALE-Dellen manual. *Figshare*, https://doi.org/10.6084/m9.figshare.3473690.v2

Collins, G. S., Wünnemann, K., Artemieva, N., & Pierazzo, E. (2013). Numerical modelling of impact processes. *in* Impact cratering: Processes and Products, 1$^{st}$ edition, edited by Osinski, G. R. and Pierazzo, E., 17, 254-270.

Davison, T. M., Collins, G. S., & Ciesla, F. J. (2010). Numerical modeling of heating in porous planetesimal collisions. *Icarus*, **208**, 468-481.

Davison, T. M., Collins, G. S., & Bland, P. A. (2016). Mesoscale modeling of impact compaction of primitive solar system solids. *The Astrophysical Journal*, 821:68.

Davison, T. M., Derrick, J. G., Collins, G. S., Bland, P. A., Rutherford, M. E., Chapman, D. J. & Eakins, D. E. (2017). Impact-induced compaction of primitive solar system solids: The need for mesoscale modelling and experiments. *Procedia Engineering*, 204, 405-412.

Duke, M. B., & Silver, L. T. (1967). Petrology of eucrites, howardites and mesosiderites. *Geochimica et Cosmochimica Acta*, *31*, 1637-1665.

Güldemeister, N., Moreau, J.-G., Kohout, T., Luther, R., & Wünnemann, K. (2022). Insight into the distribution of high-pressure shock metamorphism in Rubble-pile asteroids. *The Planetary Science Journal*, 3:198.

Güldemeister, N., Wünnemann, K., Durr, N., & Hiermaier, S. (2013). Propagation of impact-induced shock waves in porous sandstone using mesoscale modeling. *Meteoritics & Planetary Science*, 48, 115-133.

Johnson, G. R. & Cook, W. H. (1983). A constitutive model and date for metals subjected to large strains, high strain rates and high temperatures. Seventh International Symposium on Ballistics, Hague.

Japel, S., Boehler, R., & Ross, M. (2005). Melting of copper and nickel at high pressure: the role of d-electrons. *Phys. Rev. Lett.*, *95*: 167801.

Köster, W. & Franz, H. (1961). Poisson's ratio for metals and Alloys. *Metallurgical Review*, *6*: 1-56.

Kowitz, A., Güldemeister, N., Reimold, W. U., Schmidt, R. T., & Wünnemann, K. (2013). Diaplectic quartz glass and SiO2 melt experimentally generated at only 5 GPa shock pressure in porous sandstone: Laboratory observations and meso-scale numerical modeling. *Earth and Planetary Science Letters*, 384, 17-26.

Kurosawa, K., Ono, H., Niihara, T., Sakaiya, T., Kondo, T., Tomioka, N., Mikouchi, T., Genda, H., Matsuzaki, T., Kayama, M., Koike, M., Sano, Y., Murayama, M., Satake, W., & Matsui, T. (2022). Shock recovery with decaying compressive pulses: Shock effects in calcite (CaCO3) around the Hugoniot elastic limit. *Journal of Geophysical Research: Planets*, *127*, e2021JE007133. https://doi.org/10.1029/2021JE007133

Macke, R. J., Britt, D. T., & Consolmagno, G. J. (2011). Density, porosity, and magnetic susceptibility of achondritic meteorites. *Meteoritics & Planetary Science*, *46*, 311-326.

Mayne, R. G., McSween Jr, H. Y., McCoy, T. J., & Gale, A. (2009). Petrology of the unbrecciated eucrites. *Geochimica et Cosmochimica Acta*, **73**(3), 794-819.

Mittlefehldt, D. W. (2015). Asteroid (4) Vesta: I. The howardite-eucrite-diogenite (HED) clan of meteorites. *Chemie der Erde*, *75*, 155-183.

Mittlefehldt, D. W., Greenwood, R. C., Berger, E. L., Le, L., Peng, Z. X., & Ross, D. K. (2022). Eucrite-type achondrites: Petrology and oxygen isotope compositions. *Meteoritics & Planetary Science*, *57*, 484-526.

Moreau, J.-G., Kohout, T., Wunnemann, K., Holodova, P., & Haloda, J. (2019). Shock physics mesoscale modeling of shock stage 5 and 6 in ordinary and enstatite chondrites. *Icarus,* 332, 50-65.

Nakamura, A. M., Yamane, F., Okamoto, T., & Takasawa, S. (2015). Size dependence of the disruption threshold: Laboratory examination of millimeter–centimeter porous targets. *Planetary and Space Science*, 107, 45-52.

Nakazawa, S., Watanabe, S., Iijima, Y., & Kato, M. (2002). Experimental investigation of shock wave attenuation in basalt. *Icarus*, *156*, 539–550.

Ono, H., Takenouchi, A., Mikouchi, T., & Yamaguchi, A. (2019). Silica minerals in cumulate eucrites: Insights into their thermal histories. *Meteoritics & Planetary Science*, *54*, 2744-2757.

Park, R. S., Konopliv, A. S., Asmar, S. W., Bills, B. G., Gaskell, R. W., Raymond, C. A., Smith, D. E., Toplis, M. J., & Zuber, M. T. (2014). Gravity field explanation in ellipsoidal harmonic and polyhedral internal representation applied to Vesta. *Icarus*, 240, 118-132.

Pierazzo, E., Artemieva, N. A., Ivanov, B. A. (2005). Starting conditions for hydrothermal systems underneath Martiann craters: Hydrocode modeling. *in Large meteorite impacts III*, eds., Kenkmann, T., Hörz, F., and Deutsch, A.,: *Geological Society of America Special Paper, 384*, p. 443–457.

Pierazzo, E., Artemieva, N., Asphaug, E., Baldwin, E. C., Cazamias, J., Coker, R., Collins, G. S., Crawford, D. A., Davison, T., Elbeshause, D., Holsapple, K. A., Housen, K. R., Korycansky, D. G., & Wünnemann, K. (2008). Validation of numerical codes for impact and explosion cratering: Impacts on strengthless and metal targets. *Meteoritics and Planetary Science*, *43*, 1917-1938.

Poirier, J. P. (1991). *Introduction to the physics of the Earth's interior*. Cambridge University Press, New York.

Sato M., Kurosawa K., Kato S., Ushioda M., and Hasegawa S. (2021). Shock Remanent Magnetization Intensity and Stability Distributions of Single-Domain Titanomagnetite-
36

Bearing Basalt Sample Under the Pressure Range of 0.1–10 GPa. *Geophysical Research Letters*, *48*, 8.

Sekine, T., Kobayashi, T., Nishio, M., & Takahashi, E. (2008). Shock equation of state of basalt. *Earth, Planets and Space*, 60, 999-1003.

Sugita, S. & Schultz, P. H. (2003). Interactions between impact-induced vapor clouds and the ambient atmosphere: 2. Theoretical modeling. *Journal of Geophyical Research*, *108*, 5052.

Takeda, H., Miyamoto, M., Ishii, T., & Reid, A. M. (1976, April). Characterization of crust formation on a parent body of achondrites and the moon by pyroxene crystallography and chemistry. In Lunar and Planetary Science Conference Proceedings (Vol. 7, pp. 3535-3548).

Takeda, H., Mori, H., Delaney, J. S., Prinz, M., Harlow, G. E., & Ishii, T. (1983). Mineralogical comparison of Antarctic and non-Antarctic HED (howardites-eucrites-diogenites) achondrites. *Mem. Nat. Inst. Polar Res*., *30*, 181-205.

Thompson, S. L., & Lauson, H. S. (1972). Improvements in the Chart-D radiation hydrodynamic code III: Revised analytical equation of state. pp. *SC-RR-71 0714*, *119* pp., Sandia Laboratories, Albuquerque, NM.

Tillotson, J. H. (1962). Metallic equations of state for hypervelocity impact. Technical Report *GA-3216*, General Atomic Report.

Warren, P. H., Jerde, E. A., Migdisova, L. F., & Yaroshevsky, A. A. (1990). Pomozdino-an anomalous, high-MgO/FeO, yet REE-rich eucrite. In Lunar and Planetary Science Conference Proceedings (Vol. 20, pp. 281-297).

Winkler, R., Luther, R., Poelchau, M. H., Wünnemann, K., and Kenkmann, T. (2018), Subsurface deformation of experimental hypervelocity impacts in quartzite and marble targets, *Meteoritics & Planetary Science*, 53, 1733-1755.

Wünnemann, K., Collins, G. S., & Melosh, H. J. (2006). A strain-based porosity model for use in hydrocode simulations of impacts and implications for transient crater growth in porous targets. *Icarus*, *180*, 514–527.

Wünnemann, K., Collins, G. S., Osinski, G. R. (2008). Numerical modelling of impact melt production in porous rocks. *Earth and Planetary Science Letters*, *269*:530-539.

Yamaguchi, A., Shirai, N., Okamoto, C., & Ebihara, M. (2017). Petrogenesis of the EET 92023 achondrite and implications for early impact events. *Meteoritics & Planetary Science*, *52*, 709-721